\documentclass[twocolumn,amsmath,amssymb,superscriptaddress,pra]{revtex4}

\usepackage[utf8]{inputenc}
\usepackage{amsmath,amssymb,amsfonts,bm}
\usepackage{graphicx,epstopdf}  
\usepackage{color}
\usepackage{extarrows}
\usepackage{enumitem}
\usepackage{empheq}
\usepackage{graphicx}
\usepackage{float}
\usepackage{hhline}

\usepackage[colorlinks = false,
linkcolor = blue,
urlcolor  = blue,
citecolor = blue,
anchorcolor = blue, hidelinks]{hyperref}

\newcommand{\MYhref}[3][blue]{\href{#2}{\color{#1}{#3}}}%

\newcommand{\w}{\omega}

\newcommand{\ti}{\tilde}

\newcommand{\D}{\mbox{\tiny D}}

\newcommand{\E}{\mbox{\tiny E}}

\newcommand{\tS}{\mbox{\tiny S}}

\newcommand{\T}{\mbox{\tiny T}}

\newcommand{\dg}{\dagger}
\newcommand{\la}{\langle}
\newcommand{\ra}{\rangle}

\newcommand{\rra}{\rangle\rangle}

\newcommand{\Sec}[1]{Sec.\,\ref{#1}}
\newcommand{\nl}{\nonumber \\}
\newcommand{\Eq}[1]{Eq.\,(\ref{#1})}
\newcommand{\Eqs}[1]{Eqs.\,(\ref{#1})}
\newcommand{\Fig}[1]{Fig.\,\ref{#1}}

\newcommand{\RN}[1]{%
  \textup{\uppercase\expandafter{\romannumeral#1}}%
}

\newcommand{\be}{\begin{equation}}
\newcommand{\ee}{\end{equation}}
\newcommand{\bsube}{\begin{subequations}}
\newcommand{\esube}{\end{subequations}}

\usepackage{makecell}

\allowdisplaybreaks

\begin{document}

\title{
Towards Quantum Simulation of Non-Markovian Open Quantum Dynamics: 
\\
A Universal and Compact Theory 
}

\author{Xiang Li}
\thanks{Authors of equal contributions}
\affiliation{Key Laboratory of Precision and Intelligent Chemistry,
University of Science and Technology of China, Hefei 230026, China}

\author{Su-Xiang Lyu}
\thanks{Authors of equal contributions}
\affiliation{Key Laboratory of Precision and Intelligent Chemistry,
University of Science and Technology of China, Hefei 230026, China}

\author{Yao Wang}
\email{wy2010@ustc.edu.cn}
\affiliation{Hefei National Laboratory,  University of Science and Technology of China,  Hefei, Anhui 230088, China}
\affiliation{Hefei National Research Center for Physical Sciences at the Microscale, University of Science and Technology of China, Hefei, Anhui 230026, China}

\author{Rui-Xue Xu}
\affiliation{Key Laboratory of Precision and Intelligent Chemistry,
University of Science and Technology of China, Hefei 230026, China}
\affiliation{Hefei National Laboratory,  University of Science and Technology of China,  Hefei, Anhui 230088, China}
\affiliation{Hefei National Research Center for Physical Sciences at the Microscale, University of Science and Technology of China, Hefei, Anhui 230026, China}

\author{Xiao Zheng}
\email{xzheng@fudan.edu.cn}
\affiliation{Department of Chemistry, Fudan University, Shanghai 200433, China}

\author{YiJing Yan}
\email{yanyj@ustc.edu.cn}
\affiliation{Hefei National Research Center for Physical Sciences at the Microscale, University of Science and Technology of China, Hefei, Anhui 230026, China}

\date{Submitted on February 8, 2024; Resubmitted on September 2, 2024}

\begin{abstract}

Non-Markovianity, the intricate dependence of an open quantum system on its temporal evolution history, holds tremendous implications across various scientific disciplines. However, accurately characterizing the complex non-Markovian effects has posed a formidable challenge for numerical simulations. While quantum computing technologies show promise, a universal theory enabling practical quantum algorithm implementation has been elusive. We address this gap by introducing the dissipaton-embedded quantum master equation in second quantization (DQME-SQ). This exact and compact theory offers two key advantages: representability by quantum circuits and universal applicability to any Gaussian environment. We demonstrate these capabilities through digital quantum simulations of non-Markovian dissipative dynamics in both bosonic and fermionic environments. The DQME-SQ framework opens a new horizon for the efficient exploration of complex open quantum systems by leveraging the rapidly advancing quantum computing technologies.

\end{abstract}

\maketitle

\section{ Introduction}

The simulation of open quantum systems (OQSs) has attracted wide attention,
because their interactions with the surrounding environment lead to intricate dynamical behavior \cite{Wei21,Bre02, Kle09}.
OQSs often exhibit non-Markovian features, characterized by memory effects where past states influence the current state.
Understanding and characterizing non-Markovian effects are crucial for various physical processes such as quantum transport, quantum dephasing, coherent energy transfer, and precise measurement and control of quantum states
\cite{Nit06,Bro16045005,Zhe13086601,Bre16021002, Dev17015001}.
For example, in biomolecular systems such as the Fenna-Matthews-Olson complex and the B800-850 ring, the  non-Markovianity facilitates highly efficient 
energy conversion and  charge transfer 
\cite{Lee071462,Eng07782,Nal15022706,Kun22015101}.
In strongly correlated quantum dots, non-Markovian memory leads to memristive behavior that may be harnessed in novel quantum device applications \cite{Zhe13086601}.
However, accurate characterization of non-Markovian effect in complex OQSs has remained a longstanding challenge.

Among the numerous attempts \cite{Sha045053,Sto02170407,Moi13134106,Han19050601,Gar974636,Tam18030402, Tam19090402,Lam193721,Sue14150403,Zha23080901,Ke22194102,Xu22230601,Ke16024101,Luo23030316} to address this challenge, a representative method, the hierarchical equations of motion (HEOM)  method \cite{Tan89101,Tan906676,Tan20020901,Yan04216,Xu07031107}, has gained significant attention.
The HEOM method characterizes non-Markovianity through a two-time correlation function in the form of a memory kernel, which is decomposed and reorganized within a hierarchy to account for all the possible dissipation modes.
However, the exponential growth of dynamical variables presents serious challenges in computational time and data storage, making it difficult to apply HEOM and other non-Markovian quantum dissipation methods to complex OQSs within existing classical computational frameworks \cite{Tan20020901}.

In recent years, the rapid progress in quantum technologies has launched a new era of computational capabilities, enabling exploration of previously inaccessible realms of the modern science \cite{Van044,Lad107285,Bia17195,Cao1910856, Mca201,Oll20260511,Mie2325, Dal22667,Bla12277,Gro17995}.
Quantum computation holds promise for efficiently simulating OQSs. 
Several quantum computing algorithms have been developed for simulating quantum dynamics \cite{Oll20260511,Kam22010320,Mie2325,Cir23Arxiv2311_15240,Lam23Arxiv2310_12539,Zha24054101,Bar11486,Bar22043161,Li23147,Gac24013143,Sur231002},
including full quantum algorithms \cite{Sch21270503, Sch22023216, Hea21013182}  and quantum-classical hybrid algorithms based on the variational principle \cite{Sar05250503,End20010501,Yao21030307,Wan234851,Guo24_arxiv_2404_10655}.
However, simulating non-Markovian OQSs in quantum circuits remains a challenging task.
In the HEOM method, the main difficulty lies in the complex structure of the coupled equations, leading to intricate quantum circuits representing the propagator.
A novel formalism is required for the quantum simulation of non-Markovian open quantum dynamics.

This work aims to establish a universal and compact theory, the dissipaton-embedded quantum master equation in second quantization (DQME-SQ). Its unique capabilities are demonstrated through digital quantum simulations of non-Markovian dissipative dynamics in both bosonic and fermionic environments.

The remainder of this paper is organized as follows. In \Sec{thsec2} and \Sec{thsec3}, we present the construction of DQME-SQ theories for  bosonic and fermionic environments, respectively.
The quantum simulation of non-Markoivan OQSs using DQME-SQ is elaborated in \Sec{thsec4}, with numerical demonstrations in \Sec{thsec5}.
Finally, we summarize the paper in \Sec{thsum}.

\begin{figure}[th]
  \includegraphics[width=\columnwidth]{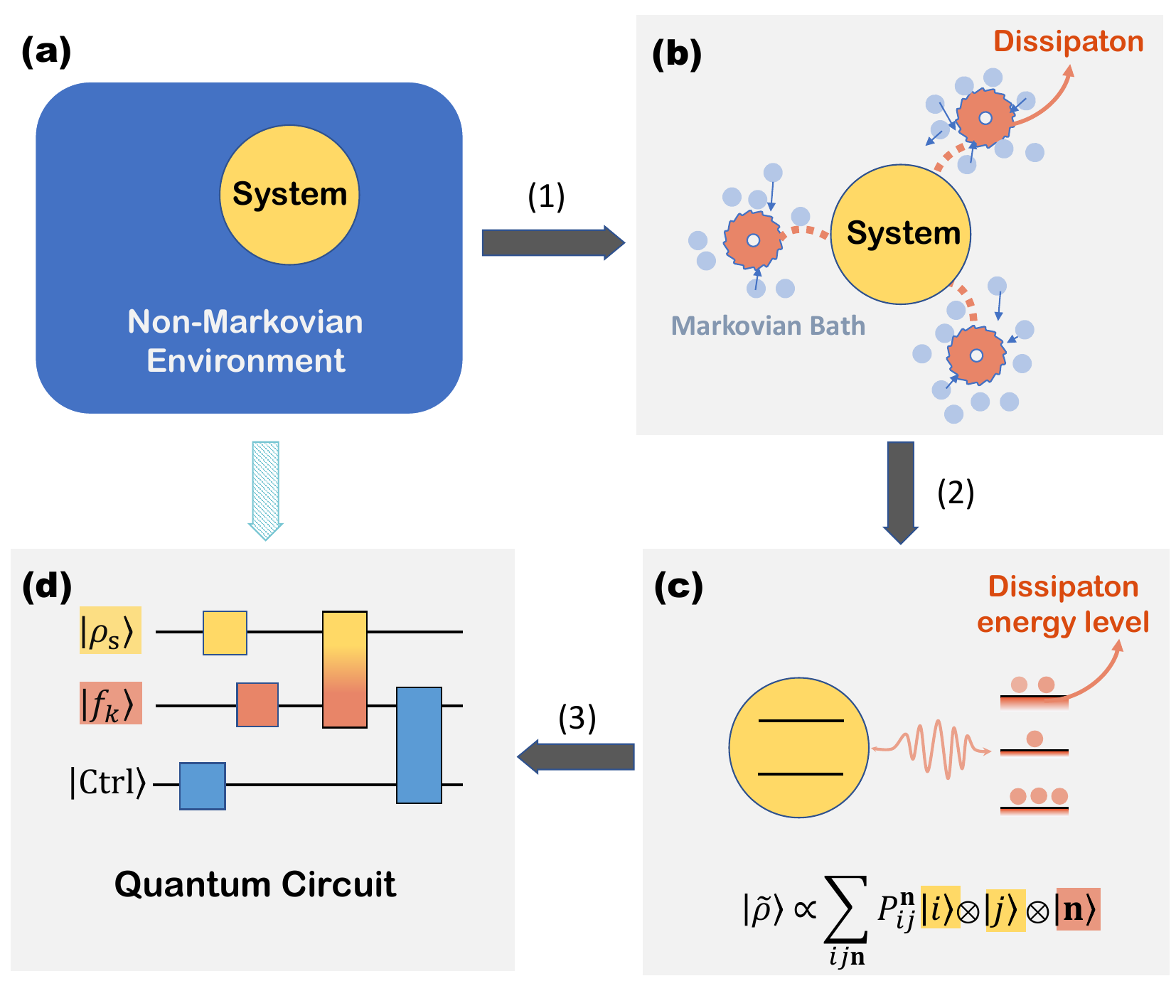}
  \caption{ 
Schematic representation of DQME-SQ theory for quantum simulation. Arrow (1) from panel (a) to (b) represents the mapping from the original environment to the individual dissipatons, statistically independent Brownian quasi-particles. In panel (c), dissipatons are second-quantized, represented by the creation and annihilation operators acting on the Fock states.
Panel (d) sketches a quantum circuit that processes the qubit states encoding the dissipatons, with $|{\rm Ctrl}\ra$ being an ancillary control bit for time propagation.
As a result, the DQME-SQ enables quantum simulation of non-Markovian open quantum dynamics.}\label{fig1}
\end{figure}

\section{Bosonic DQME-SQ theory}\label{thsec2}

\subsection{Construction of bosonic DQME-SQ}\label{thsec2_1}

In our previous work \cite{Li23214110},
we had established the dissipaton-embedded quantum master equation (DQME).
In this theory, the non-Markovian memory content of the environment is effectively represented by a set of statistically independent generalized Brownian quasi-particles, referred to as \emph{dissipatons} \cite{Yan14054105}. 
Depicted in \Fig{fig1}(a) is the total composite system, consisting of the system of primary interest and its non-Markovian environment.
In \Fig{fig1}(b), the environment is represented by a series of dissipatons (red plates) that interact with the system.
Each dissipaton behaves as a Brownian particle moving within a Markovian bath (light blue dots). 
The non-Markovian memory of the environment is manifested through the characteristic energies and lifetimes of the dissipatons. 
These dissipatons can interact strongly with the system, and are thus regarded as being embedded within the system. 
In the first quantization formulation, the DQME is a time-local Fokker-Planck-like equation \cite{Li23214110}.
Taking the bosonic open quantum system as the example, the temporal evolution of density distribution of the 
dissipaton-embedded system,  $\rho(x_1, x_2,\cdots)$, is governed by the DQME, where $x_1, x_2,\cdots$ are the coordinates of dissipatons. However, since the coordinates $\{x_k\}$ are \emph{continuous} variables, it is difficult to encode the DQME for quantum simulation purposes.

To overcome this difficulty, we notice that the quantum dissipation process can be fully characterized by a combination of generation and annihilation events of  dissipatons.
Such characterization corresponds to a second quantization procedure.
In Fig.~1(c), each short black bar corresponds to an energy level capable of accommodating any number of dissipatons, while the broadening of the level (represented by red shade) indicates the associated lifetime of dissipatons. 
A dissipaton configuration is thus established by specifying the occupancy on all the dissipaton levels.

After applying the second quantization, the resulting DQME-SQ governs the transitions between different dissipaton configurations.
In principle, the non-Markovianity is exactly accounted for if all the possible dissipaton configurations are explicitly included in the open quantum dynamics. This is analogous to the electronic structure theory where full configuration interaction treatment yields exact many-body wavefunction.

The construction of DQME-SQ will first be exemplified with an OQS with a bosonic environment.
In the following, we set $\hbar=1$ and $\beta \equiv 1/(k_{\rm B}T)$, where $k_{\rm B}$ is the Boltzmann constant and $T$ the environmental temperature.
The total Hamiltonian has the generic form of
$
H_{\T}=H_{\tS}+H_{\E}+H_{\tS\E}
$. 
Here,  $H_{\tS}$ and $H_{\E}$ are the system and environment Hamiltonian, respectively.
For brevity, the system--environment coupling is set as 
$H_{\tS\E}=\hat Q \hat F$, 
where $\hat Q$ is a system operator, and the environment
operator $\hat F$ incorporates the system--environment coupling strength. %
The non-Markovian memory of a Gaussian environment is fully characterized by
the hybridization  correlation function,
$
C(t)=\la \hat F(t)  \hat F(0)\ra_{\E}
$.
Here, $ \hat F(t)\equiv e^{iH_{\E}t} \hat Fe^{-iH_{\E}t}$, 
and $\la\hat O\ra_{\E}
\equiv {\rm Tr}_{\E}(\hat O e^{-\beta H_{\E}})/ {\rm Tr}_{\E}(e^{-\beta H_{\E}})$ 
is the ensemble average at the equilibrium state of bare environment, where $\hat O$ is an arbitrary operator.

The DQME-SQ formalism starts with the 
dissipaton decomposition (DD) [cf.\ Arrow\,(1) in \Fig{fig1}],
which should preserve
 the correlation function 
$C(t)$. The DD maps the  environment hybridization operator $\hat F$ onto $\{\hat f_k\}$, the coordinate operators of dissipatons \cite{Yan14054105, Wan22170901},
\begin{equation} \label{map1}
\hat F \stackrel{\text{DD}}{\longmapsto}\sum_{k=1}^{K}\hat f_k.
\end{equation}
For individual dissipatons, we set $\la \hat f_k(t) \hat f_{k'}(0)\ra_{\D}=\delta_{kk'}c_k(t)$, 
where $\la \hat O\ra_{\D}$ denotes the expectation value taken at the dissipaton vacuum state \cite{Wan22044102}.
It is important that the non-Markovian memory of the original bosonic environment is fully recovered through 
\begin{equation} \label{condition}
C(t)=\sum_{k,k'=1}^{K} \la \hat f_k(t)\hat f_{k'}(0)\ra_{\D}=\sum^K_{k=1}c_k(t).
\end{equation}

In this work, we adopt the function form of $c_k(t)=\eta_k e^{-\gamma_k t}$, which leads to an exponential decomposition of $C(t)$.
Numerically, such a decomposition can be accurately and routinely realized using the
Prony decomposition algorithm \cite{Che22221102} or other algorithms \cite{Xu22230601,Tak24204105}. The resulting DD protocol is universal for arbitrary spectral density of the environment and for any temperature \cite{Che22221102}.
%
It has been proved that the exponents $\{\gamma_k\}$ are either real numbers or emerge as complex conjugate pairs, and the coefficients $\{\eta_k\}$ generally take complex values \cite{Yan14054105,Yan16110306}.
Therefore, we may determine $\bar k$ by the pairwise equality $\gamma_{\bar k}=\gamma_{k}^{\ast}$,
and the imaginary and real parts of $\gamma_k$ correspond to the characteristic energy  and decaying rate
of a dissipaton, respectively.

%

The coordinate operators of the dissipatons are further second-quantized by the transformation
\begin{equation} \label{map2}
 \hat f_k\stackrel{\text{SQ}}{\longmapsto}\zeta_k \, \big(\hat b_k^{-}+\hat b_k^{+} \big).
\end{equation}
Here, the coefficients are chosen as $\zeta_k= \sqrt{(\eta_k+\eta^*_{\bar k})/2}$. $\{\hat 
 b_{k}^{+}\}$ and $\{\hat  b_k^{-}\}$ are bosonic creation and annihilation operators, respectively. 
They satisfy the relation $[\hat b^-_k,\hat b^+_{k'}]=\delta_{kk'}$.
For each dissipaton, the Fock space ${\cal H}_{f_k}$ is then generated,
and one can thus establish the DQME-SQ for 
the reduced density tensor (RDT) of the dissipaton-embedded system, denoted by $\ti \rho(t)$. 
Here, $\ti\rho$ spans the tensor product space of the original system and the dissipaton Fock space, i.e.,
$
  \tilde \rho \in {\cal H}_{\tS}\otimes{\cal H}^{\ast}_{\tS}\otimes {\cal H}_{\D}
$.
The DQME-SQ for $\ti \rho(t)$ reads
\begin{align}\label{dqme-sq}
\dot{\tilde \rho}(t)\!=\!&-i[H_{\tS},\tilde\rho]-\sum_k \gamma_k \hat N_k \tilde \rho 
\nl
\!&-i\sum_k \!\zeta_k(\hat b_k^+ + \hat b_k^-)[\hat Q,\tilde\rho]+\!\sum_k \xi_k \hat b_k^+  \{\hat Q,\tilde\rho\}.
\end{align}
Here, $\{\cdot\,,\,\cdot\}$ denotes the anti-commutator, $\hat N_k\equiv\hat b_k^+ \hat b_k^-$ and $
  \xi_k =  (\eta_k -\eta^*_{\bar k})/(2i\zeta_k) $.
On the right-hand side of \Eq{dqme-sq}, the second term describes the decay of the dissipatons, and the last two terms account for the interactions between the system and the dissipatons.

The DQME-SQ of \Eq{dqme-sq} yields exact non-Markovian dynamics for any Gaussian environment, provided that the hybridization correlation function of the environment can be decomposed into a sum of exponential functions. 
This exactness will be validated in \Sec{thsec2_2} by establishing a formal equivalence between \Eq{dqme-sq} and the bosonic HEOM. 
Moreover, it will also be demonstrated in \Sec{thsec2_3} that \Eq{dqme-sq} is universally applicable to general OQSs, where the system contains multiple degrees of freedom and is coupled to more than one environment.

The pseudomode theory \cite{Gar974636,Tam18030402, Tam19090402,Lam193721,Cir23Arxiv2311_15240,Lam23Arxiv2310_12539} has been established based on an exponential decomposition of the environment correlation function $C(t)$, leading to a Lindblad equation capable of capturing the exact non-Markovian open quantum dynamics, subject to the condition that the pre-exponential coefficients $\{\eta_k\}$ are real numbers, as implied by Eq.~(18) in Ref.~[\onlinecite{Tam18030402}] and Eq.~(35) in SM of Ref.~[\onlinecite{Lam193721}]. 
Under this condition, an explicit relationship between \Eq{dqme-sq} and the Lindblad equation in the pseudomode theory can be established; see Appendix \ref{thapp1}. However, it is worth noting that \Eq{dqme-sq} is applicable to more general situations with complex-valued $\{\eta_k\}$. Moreover, the Hilbert space associated with \Eq{dqme-sq} is more compact than that of the Lindblad equation in the pseudomode theory, which is given by 
$ {\cal H}_{\tS}\otimes{\cal H}^{\ast}_{\tS}\otimes {\cal H}_{\D}\otimes {\cal H}_{\D}^{\ast} $.

\subsection{Exactness of bosonic DQME-SQ}\label{thsec2_2}

To verify the exactness of the bosonic DQME-SQ theory, in the following we will give a detailed derivation of \Eq{dqme-sq} by starting from the bosonic HEOM formalism, which is known to be an exact approach \cite{Tan20020901}.  
As explained in \Sec{thsec2_1}, we start by expanding the hybridization correlation function of the bosonic environment as a linear combination of exponential functions, 
\be \label{exp}
C(t)=\sum_{k}\eta_k e^{-\gamma_k t},
\ee
where $\{\eta_k\}$ and $\{\gamma_k\}$ are complex numbers in general.

The exact bosonic HEOM can then be constructed as \cite{Yan16110306, Tan20020901, Wan22170901}
\begin{align} \label{DEOMB}
\dot\rho^{(n)}_{\bf n}&=\!
  -i\big[{H}_{\tS},\rho^{(n)}_{\bf n}\big]\!-\sum_k n_k \gamma_{k}\rho^{(n)}_{\bf n}  
   -i\sum_{k}\!\big[\hat Q,\rho^{(n+1)}_{{\bf n}_{k}^+}\big]
\!
\nl & \quad
-i\sum_{k}\!n_{k}\Big(\eta_{k}\hat Q\rho^{(n-1)}_{{\bf n}_{k}^-}
    \!-\eta_{\bar k}^{\ast}\rho^{(n-1)}_{{\bf n}_{k}^-}\hat Q\Big).
 \end{align}
The dynamical variables in \Eq{DEOMB} are
$\rho^{(n)}_{\bf n}(t)\equiv \rho^{(n)}_{n_1\cdots n_K}(t)$, where $n=n_1+\cdots+n_{K}$, with $n_k\geq 0$.
In other words, 
each variable $\rho^{(n)}_{\bf n}(t)$ is associated with
an ordered set of indexes, ${\bf n}\equiv \{n_1\cdots n_K\}$.
Denote also ${\bf n}^{\pm}_{k}$ that differs from ${\bf n}$ only
at the specified $k$-occupation number
$n_{k}$ by $\pm 1$.
The reduced density operator of the system is  
$\rho_{\tS}(t)\equiv \rho_{\bf 0}^{(0)}(t)\equiv \rho_{0\cdots 0}^{(0)}(t)$.

To proceed, we make use of two sets of complex-valued parameters, $\{\zeta_k\}$ and   
$\{\xi_k\}$, given below \Eq{map2} and \Eq{dqme-sq}, respectively. 
Define dimensionless variables
\be \label{barrho}
\bar {\rho}_{\bf n}^{(n)}(t)\equiv \rho_{\bf n}^{(n)}(t)/ \prod_k \big(\zeta^{n_k}_k\sqrt{n_k !} \big).
\ee
Equation (\ref{DEOMB}) is then recast into \cite{Li23214110}
\begin{align}\label{DEOM2}
\dot{\bar{\rho}}^{(n)}_{\bf{n}}=&-i[H_{\tS},\bar{\rho}^{(n)}_{{\bf n}}] -\sum_k {n_k}\gamma_k \bar{\rho}^{(n)}_{{\bf n}}
\nl &
-i\sum_{k}\zeta_k \hat{Q}^{\times}\Big[\sqrt{n_k +1}\bar{\rho}^{(n+1)}_{{\bf n}^+_k}+\sqrt{n_k}\bar{\rho}^{(n-1)}_{{\bf n}^-_k} \Big]
\nl
&+\sum_{k} \sqrt{ n_k} \xi_{k}\hat{Q}^{\diamond}\bar{\rho}^{(n-1)}_{{\bf n}^-_k}.
\end{align}
Here, we denote the superoperators $\hat A^{\times}\hat O\equiv \hat A\hat O-\hat O \hat A$ and  $\hat A^{\diamond}\hat O\equiv \hat A\hat O+\hat O \hat A$ for any operators $\hat A$ and $\hat O$.

To derive the bosonic DQME-SQ,  we take the RDT
of the dissipaton-embedded system $\ti \rho(t)$ as a state vector in the space $ \mathcal{H}_{\tS}\otimes \mathcal{H}_{\tS}^* \otimes {\cal H}_{\D}$, i.e., the reduced system space ${\cal H}_{\tS}\otimes{\cal H}^{\ast}_{\tS}$ further dilated by the dissipaton Fock space ${\cal H}_{\D}=\otimes_k {\cal H}_{f_k}$.
It reads
\bsube
\begin{equation}\label{defB}
\tilde{\rho}(t)\equiv \sum_{\bf n} \bar {\rho}_{\bf n}^{(n)}(t)\otimes |{\bf n}\ra
\end{equation}
with
\begin{equation}
|{\bf n} \ra=|n_1\ra\otimes |n_2\ra\otimes \cdots \otimes |n_K\ra.
 \end{equation}
 \esube
Here, $\{\bar{\rho}_{\bf n}^{(n)}(t)\}$ are defined in  \Eq{barrho}, and $|n_k\ra \in \mathcal{H}_{f_k}$ presents the occupancy of the $k$th dissipaton level, which is generated by bosonic creation and annihilation operators  $\{\hat b_k^{\pm}\}$  following the standard bosonic commutation algebra 
\begin{align}
[\hat b^-_k,\hat 
b^+_{k'}]=\delta_{kk'} \quad \text{and} \quad
[\hat b^\pm_k,\hat 
 b^\pm_{k'}]=0,
\end{align}
as well as the relations,
\bsube
\begin{align}
  &\hat  b_k^- |{\bf 
 n}\ra=\sqrt{n_k}|{\bf 
n}^{-}_{k}\ra,
\\
 & \hat  b_k^+ |{\bf 
 n}\ra=\sqrt{n_k+1}|{\bf 
n}^{+}_{k}\ra.
\end{align}
\esube
Combining \Eq{DEOM2} and \Eq{defB}, along with the relations
\begin{align*}
&\hat b^-_k\tilde{\rho}=\sum_{\bf n}\bar{\rho}^{(n+1)}_{{\bf n}^+_k}\otimes\hat  b^-_k|{\bf n}^+_k\ra
=\sum_{\bf n}\sqrt{n_k +1}\bar{\rho}^{(n+1)}_{{\bf n}^+_k} \otimes|{\bf n}\ra,
\\
&\hat b^+_k\tilde{\rho}=\sum_{\bf n}\bar{\rho}^{(n-1)}_{{\bf n}^-_k}\otimes \hat b^+_k|{\bf n}^-_k \ra
=\sum_{\bf n}\sqrt{n_k }\bar{\rho}^{(n-1)}_{{\bf n}^-_k}\otimes|{\bf n}\ra,
\end{align*}
and
\begin{align*}
&\hat N_k \tilde{\rho}=\sum_{\bf n} \bar{\rho}^{(n)}_{{\bf n}} \otimes  \hat  b^+_k \hat b^-_k  |{\bf n}\ra 
=\sum_{\bf n} n_k  \bar{\rho}^{(n)}_{{\bf n}} \otimes |{\bf n}\ra,
\end{align*}
we then obtain the dynamical equation for  the RDT $\ti  \rho(t)$, which  is exactly the  bosonic DQME-SQ of \Eq{dqme-sq}.

\subsection{Universality and compactness} \label{thsec2_3}

It is straightforward to extend the DQME-SQ of \Eq{dqme-sq} to OQSs containing multiple degrees of freedom that are coupled to more than one environment. 
For instance, consider the case of $H_{\tS\E}=\sum_{\alpha u}\hat Q_u\hat F_{\alpha u}$,  where $\alpha$ and $u$ label the specific environment and degree of freedom, respectively.
The exponential decomposition reads
$C_{\alpha u}(t)\equiv \la \hat F_{\alpha u}(t)\hat F_{\alpha u}(0)\ra_{\E}=\sum_{k}\eta_{\alpha uk} e^{-\gamma_{\alpha uk} t}$, and the corresponding
DQME-SQ is
\begin{align}
\dot{\tilde \rho}&=-i[H_{\tS},\tilde\rho]-\sum_{\alpha uk} \gamma_{\alpha uk} \hat  b_{\alpha uk}^+ \hat 
 b_{\alpha uk}^- \tilde \rho +\sum_{\alpha uk} \xi_{\alpha uk} \hat  b_{\alpha uk}^+  \{\hat{Q}_u,\tilde\rho\}
\nl & \quad
-i\sum_{\alpha uk} \zeta_{\alpha uk}(\hat  b_{\alpha uk}^+ +\hat  b_{\alpha uk}^-)[\hat{Q}_u,\tilde\rho].
\end{align}
Here, $\zeta_{\alpha uk}\equiv \sqrt{(\eta_{\alpha uk}+\eta^*_{\alpha u\bar k})/2}$ and 
$
\xi_{\alpha uk}\equiv (\eta_{\alpha uk}-\eta^*_{\alpha u\bar k})/(2i\zeta_{\alpha uk})
$, and $\ti \rho$ is represented as a state vector in the Hilbert space 
$\mathcal{H}_{\tS}\otimes \mathcal{H}_{\tS}^* \otimes {\cal H}_{\D}$, i.e., the system space ${\cal H}_{\tS}\otimes{\cal H}^{\ast}_{\tS}$ dilated by the dissipaton Fock space ${\cal H}_{\D}=\otimes_{\alpha uk} {\cal H}_{f_{\alpha uk}}$.

As evidenced in Appendix~\ref{thapp1}, the bosonic DQME-SQ fully captures the non-Markovian open quantum dynamics within a smaller Hilbert space compared to the pseudomode theory, without any approximation.


\section{Fermionc DQME-SQ theory}\label{thsec3}

\subsection{Construction of fermionic DQME-SQ}\label{thsec3_1}

For fermionic environments, the DQME-SQ theory can be established in a similar way. 
For a single-level system, we consider a common bilinear form of system-environment coupling, represented by
$H_{\tS\E}=\hat c^{+} \hat F^{-} + \hat F^{+} \hat c^{-}$, which describes the exchange of fermionic particles between the system and environment.
Here,  $\hat c^{-}\equiv \hat c$ ($\hat c^{+}\equiv \hat c^{\dg}$) denotes the electron annihilation (creation) operator in the system state.
The DD for the environment hybridization operators reads
\begin{equation} \label{map1_fermi}
\hat F^{\pm} \stackrel{\text{DD}}{\longmapsto}\sum_{k=1}^{K}\hat f_{k}^{\pm}.
\end{equation}
Here, $\la \hat f_k^\sigma (t) \hat f_{k'}^{\bar \sigma}(0) \ra_{\D}=\delta_{kk'}\eta_{k}^{\sigma}e^{-\gamma_{k}^{\sigma}t}$
with $\sigma=\pm$ and $\bar \sigma=-\sigma$.
Such a DD protocol completely recovers the hybridization correlation functions. 
The fermionic dissipaton Fock space is
$
{\cal H}_{\D}=\otimes_k ({\cal H}_{f^{-}_k}\otimes {\cal H}_{f^{+}_k})
$, which involves both left and right actions of dissipaton creation and annihilation operators. 
Thus, the second quantization of fermionic dissipatons is realized through the following mappings.
\begin{subequations}
\begin{align}
&\hat f_{k}^{+}\hat O\stackrel{\text{SQ}}{\longmapsto} \zeta_{k}^{+} \hat b_k^{+}\hat O+\xi_{k}^{-}\hat O\hat b_k^{+},
\\
&\hat O\hat f_{k}^{+}\stackrel{\text{SQ}}{\longmapsto} \zeta_{k}^{+} \hat b_k^{+}\hat O+\xi_{k}^{+\ast}\hat O\hat b_k^{+},
 \\
&\hat f_{k}^{-}\hat O\stackrel{\text{SQ}}{\longmapsto}  \zeta_{k}^{-} \hat O\hat b_k^{-}+\xi_{k}^{+}\hat b_k^{-}\hat O,
\\
&\hat O\hat f_{k}^{-}\stackrel{\text{SQ}}{\longmapsto} \zeta_{k}^{-} \hat O\hat b_k^{-}+\xi_{k}^{-\ast}\hat b_k^{-}\hat O,
\end{align}
\end{subequations}
where 
$\{\hat b_{k}^{+}\}$ and $\{\hat  b_k^{-}\}$ are fermionic creation and annihilation operators that satisfy $\{\hat b^-_k,\hat b^+_{k'}\}=\delta_{kk'}$,
and the coefficients are chosen as
$\zeta_{k}^{\sigma}=(\eta_{k}^{\sigma}\eta_{k}^{\bar\sigma\ast})^{\frac{1}{4}}$ and $\xi_{k}^{\sigma}=\eta_{k}^{\sigma}/\zeta_{k}^{\sigma}$.
The fermionic DQME-SQ reads
\begin{align}\label{fermion}
\!\!\!\!\dot{\tilde{\rho}} =& -i[H_{\tS},\tilde{\rho}]-\sum_k \big(\gamma^-_k \hat N_k \tilde{\rho}+ \gamma^+_k \tilde{\rho} \hat N_k\big) 
\nl
&-i\sum_{k}\!\Big[\zeta_k^{-}\big(\hat c^+ \hat b^-_k\tilde{\rho} - \!\hat b_k^-\tilde{\rho}\hat c^+)+ \zeta_k^{+}(\hat c^-\tilde \rho \hat b^+_k  \!- \!\tilde\rho \hat b^+_k \hat c^-\!\big)
\nl &
+\!\xi^+_k \hat c^+ \tilde\rho \hat b^-_k
\!-\! \xi^{+*}_k \hat b^+_k\tilde\rho \hat c^- \!\!-\!\xi^-_k \hat c^- \hat b^+_k \tilde\rho+\!\xi^{-*}_k  \tilde\rho \hat b^-_k \hat c^+\!
\Big],
\end{align}
where $\hat N_k\equiv \hat b_k^{+}\hat b_k^{-}$ is the fermionic dissipaton number operator. 
The detailed derivation of \Eq{fermion} and its formal equivalence to the fermionic HEOM are presented in \Sec{thsec3_2}. Similar to the bosonic DQME-SQ, \Eq{fermion} can be straightforwardly extended to more complex OQSs; see \Sec{thsec3_3}.

\subsection{Exactness of fermionic DQME-SQ }\label{thsec3_2}

Similar to the bosonic case, to verify the exactness of the fermionic DQME-SQ theory, we shall give a detailed derivation of \Eq{fermion} by starting from the fermionic HEOM formalism, which is known to be exact. 
For the quantum impurity model described above, the fermionic HEOM reads \cite{Jin08234703,Yan16110306}
\begin{align} \label{fermionicHEOM}
\dot{\rho}_{{\bm j}}^{(n)}&=-\bigg(i{\cal L}_{\tS}+\sum_{r=1}^{n}\gamma_{j_r}\bigg)\rho_{{\bm j}}^{(n)}-i\sum_{\ell}{\cal A}^{\bar\sigma}\rho_{{\bm j}\ell}^{(n+1)}
\nl & \quad
-i\sum_{r=1}^{n}(-)^{n-r}{\cal C}_{j_r}\rho_{{\bm j}_r^{-}}^{(n-1)}.
\end{align}
Here,  ${\cal L}_{\tS} \cdot \equiv [H_{\tS},\,\cdot\,]$, $j\equiv (k,\sigma)$, 
and ${\cal A}^{\bar\sigma}$ and ${\cal C}_j\equiv {\cal C}_{k}^{\sigma}$ are Grassmann superoperators defined by
\be 
{\cal A}^{\sigma}\hat O^{(n)}\equiv \hat c^{\sigma}\hat O^{(n)}+(-)^{n} \hat O^{(n)}\hat c^{\sigma} 
\ee
and 
\be 
{\cal C}^{\sigma}_{k}\hat O^{(n)}\equiv \eta_{k}^{\sigma}\hat c^{\sigma}\hat O^{(n)}-(-)^{n}\eta_{k}^{\bar\sigma\ast}\hat O^{(n)}\hat c^{\sigma}.
\ee
Here, $\hat O^{(n)}$ denotes 
$\rho^{(2m)}$ or $\rho^{(2m+1)}$ with even  or odd fermionic parity, respectively.
In the occupation number representation, we split the indices $\sigma=+$ and $-$, and rewrite the reduced density operator  and auxiliary density operators as
\be
\rho_{\bm j}^{(n)}\longmapsto\rho^{}_{{\bf n}{\bf m}}\equiv  \rho_{n_1n_2\cdots n_K;m_1m_2\cdots m_K},
\ee
where $n_k=1$ if $(k,+)\in {\bm j}$; and otherwise $n_k=0$. Similarly, $m_k=1$ when $(k,-)\in {\bm j}$; and otherwise $m_k=0$.
Consequently, the HEOM of \Eq{fermionicHEOM} can be recast as 
\begin{align}\label{fermionicHEOM2}
&\dot{\rho}^{}_{{\bf n}{\bf m}}=-\big(i{\cal L}_{\tS}+\gamma_{{\bf n}{\bf m}}\big)\rho^{}_{{\bf n}{\bf m}}
\nl & \quad
-i\sum_{k} \Big[(-)^{M+N-\theta^{+}_k}\hat c^{-}\rho^{}_{{\bf n}_{k}^{+}{\bf m}}
-(-)^{\theta^{+}_k}\rho^{}_{{\bf n}_{k}^{+}{\bf m}}\hat c^{-}
\nl & \quad
+(-)^{M-\theta^{-}_k}
\hat c^{+}\rho^{}_{{\bf n}{\bf m}_{k}^{+}}
-(-)^{N+\theta^{-}_k}
\rho^{}_{{\bf n}{\bf m}_{k}^{+}}\hat c^{+}
\nl &\quad
+(-)^{M+N-\theta^{+}_{k}}\eta_{k}^{+}\hat c^{+}\rho^{}_{{\bf n}_{k}^{-}{\bf m}}
-(-)^{\theta^{+}_{k}-1}(\eta_{k}^{-})^{\ast}\rho^{}_{{\bf n}_{k}^{-}{\bf m}}\hat c^{+}
\nl & \quad
+(-)^{M-\theta^{-}_{k}}\eta_{k}^{-}\hat c^{-}\rho^{}_{{\bf n}{\bf m}_{k}^{-}}-(-)^{N-1+\theta^{-}_{k}}(\eta_{k}^{+})^{\ast}\rho^{}_{{\bf n}{\bf m}_{k}^{-}}\hat c^{-} \Big].
\end{align}
Here, we denote $\gamma_{{\bf n}{\bf m}}\equiv \sum_{k=1}(n_k\gamma_{k}^{+}+m_k\gamma_{k}^{-}) $, $\theta^{+}_k=\sum_{l=1}^{k} n_l$ and $\theta^{-}_k=\sum_{l=1}^{k} m_l$, with $N=\theta_{K}^{+}$ and $M=\theta_{K}^{-}$.
%
In deriving \Eq{fermionicHEOM2}, we have used the relations
\begin{align*}
\rho_{\bm j;(k,+)}^{(n+1)}&\longmapsto   (-)^{\theta^{-}_K+\theta^{+}_K-\theta^{+}_k}\rho_{n_1n_2\cdots n_{k}+1 \cdots n_K;m_1m_2\cdots m_K}
\nl & \quad \quad
= (-)^{M+N-\theta^{+}_k}\rho^{}_{{\bf n}_{k}^{+}{\bf m}},
\end{align*}
and
\begin{align*}
\rho_{\bm j;(k,-)}^{(n+1)}&\longmapsto  (-)^{\theta^{-}_K-\theta^{-}_k}
\rho_{n_1n_2\cdots n_K;m_1m_2\cdots m_{k}+1\cdots m_K}
\nl & \quad \quad
=
(-)^{M-\theta^{-}_k}
\rho^{}_{{\bf n}{\bf m}_{k}^{+}}.
\end{align*}

To obtain \Eq{fermion}, we introduce the fermionic dissipaton creation and annihilation operators, $\{\hat{b}^\pm_k\}$, which satisfy the standard fermionic anti-commutation relations. They also anti-commute with  $\{\hat{c}^\pm\}$.
The RDT $\tilde \rho(t)$ is defined as
\begin{equation}
 \tilde\rho(t)=\sum_{\bf mn}  \bar{\rho}^{}_{\bf nm}(t) |{\bf m} {\bf n} \rra. \label{rhot}
\end{equation} 
Here, 
\be
\bar{\rho}^{}_{\bf nm}(t)=\prod_{kj} (\zeta_{k}^-)^{-m_k}(\zeta_{j}^+)^{-n_{j}} \rho^{}_{\bf nm}(t),  \label{eqn:brho-1}
\ee
where
$\rho^{}_{\bf nm}(t)$ follow \Eq{fermionicHEOM2}, $\zeta_{k}^{\sigma}=(\eta_{k}^{\sigma}\eta_{k}^{\bar\sigma\ast})^{\frac{1}{4}}$ with $\sigma=\pm$, $|{\bf mn}\rra=|{\bf m}\ra\la {\bf n}|$,
and
\begin{align}
&|{\bf m}\ra\equiv (\hat{b}^+_1)^{m_1}\cdots (\hat{b}^+_K)^{m_K} |{\bf 0}\ra
\\
&|{\bf n}\ra\equiv (\hat{b}^+_K)^{n_K}\cdots (\hat{b}^+_1)^{n_1}|{\bf 0}\ra.
\end{align}
By substituting \Eqs{fermionicHEOM2} and \eqref{eqn:brho-1} into \Eq{rhot}, we arrive at the fermionic DQME-SQ of \Eq{fermion}.

\subsection{Universality and compactness}\label{thsec3_3}

It is straightforward to extend \Eq{fermion} to general OQSs containing multiple degrees of freedom that are coupled to more than one environment,  without any loss of exactness. 
Consider the scenario of $H_{\tS\E}=\sum_u \hat c_u^{+} \hat F_{\alpha u}^{-} + \hat F_{\alpha u}^{+} \hat c_u^{-}$, where $\alpha$ and $u$ label the environments and the system's degrees of freedom, respectively. The fermionic DQME-SQ reads
\begin{align}\label{fermion_multi}
\dot{\tilde{\rho}}& = -i[H_{\tS},\tilde{\rho}]-\sum_{\alpha uk} \big(\gamma^-_{\alpha uk} \hat N_{\alpha uk} \tilde{\rho}+ \gamma^+_{\alpha uk} \tilde{\rho} \hat N_{\alpha uk}\big) 
\nl & \quad
-i\sum_{\alpha uk}\! \Big[\zeta_{\alpha uk}^{-}\big(\hat c_u^+ \hat b^-_{\alpha uk}\tilde{\rho} - \!\hat b_{\alpha uk}^-\tilde{\rho}\hat c_u^+)
\nl & \quad\quad\quad\ 
+\zeta_{\alpha uk}^{+}(\hat c_u^-\tilde \rho \hat b^+_{\alpha uk}  \!- \!\tilde\rho \hat b^+_{\alpha uk} \hat c_u^-\!\big)
\nl & \quad\quad\quad\  
+\xi^+_{\alpha uk} \hat c_{u}^+ \tilde\rho \hat b^-_{\alpha uk}
- \xi^{+*}_{\alpha uk} \hat b^+_{\alpha uk}\tilde\rho \hat c_{u}^-
\nl & \quad\quad\quad\  
-\xi^-_{\alpha uk} \hat c_{u}^- \hat b^+_{\alpha uk} \tilde\rho+\xi^{-*}_{\alpha uk}  \tilde\rho
 \hat b^-_{\alpha uk} \hat c_{u}^+
\Big],
\end{align}
where the coefficients are chosen as
$\zeta_{\alpha uk}^{\sigma}=(\eta_{\alpha uk}^{\sigma}\eta_{\alpha uk}^{\bar\sigma\ast})^{\frac{1}{4}}$ and $\xi_{\alpha uk}^{\sigma}=\eta_{\alpha uk}^{\sigma}/\zeta_{\alpha uk}^{\sigma}$.

The pseudofermion theory \cite{Cir23033011} presents a fermionic variant of the pseudomode approach. This theory utilizes a Hilbert space with a dimension identical to that associated with \Eq{fermion}, given by
$
\otimes_k ({\cal H}_{f^{-}_k}\otimes {\cal H}_{f^{+}_k})
$.
While it is speculated that a connection between the pseudofermion and DQME-SQ theories exists,
directly establishing this relationship is challenging.
The difficulty arises from the fact that the pseudofermion Lindblad equation involves projection operators associated with the even and odd fermionic parity spaces, as shown in Eqs.\,(16)-(17) of Ref.~\cite{Cir23033011}. In contrast, this parity-based structure is not required by the DQME-SQ theory.

The compactness of the fermionic DQME-SQ arises from the direct construction of fermionic dissipatons based on the exponential decomposition of hybridization correlation functions, which generally results in complex-valued coefficients and exponents. On the contrary, the pseudofermion theory  proposed in Ref.~\cite{Cir23033011} requires an additional post-processing procedure to accommodate the complex-valued coefficients from the exponential decomposition.
The post-processing procedure typically increases the number of pseudomodes required, which inevitably makes the numerical simulation more expensive. 
Such a post-processing procedure is not needed in the implementation of the DQME-SQ approach.



The fermionic DQME-SQ approach can be directly applied to investigate properties related to the environment hybridization operators, ${\hat F_u^{\pm}}$.
For example, for the electric  current defined as
$\hat{I}=-i\sum_u\big(\hat{c}_u^+\hat{F}_{u}^--\hat{F}_{ u}^+\hat{c}_u^-\big)$, its expectation value is evaluated using the  DQME-SQ  theory as
\be 
I(t) = i\sum_{uk} {\rm tr_{\tS}}\big[c_{u}^- \la {\bm 0}| \tilde{\rho}(t)|{\bm 0}^+_{uk}\ra- \la {\bm 0}^+_{uk}| \tilde{\rho}(t)|{\bm 0}\ra c_u^+\big].
\ee

\begin{figure}
  \includegraphics[width=\columnwidth]{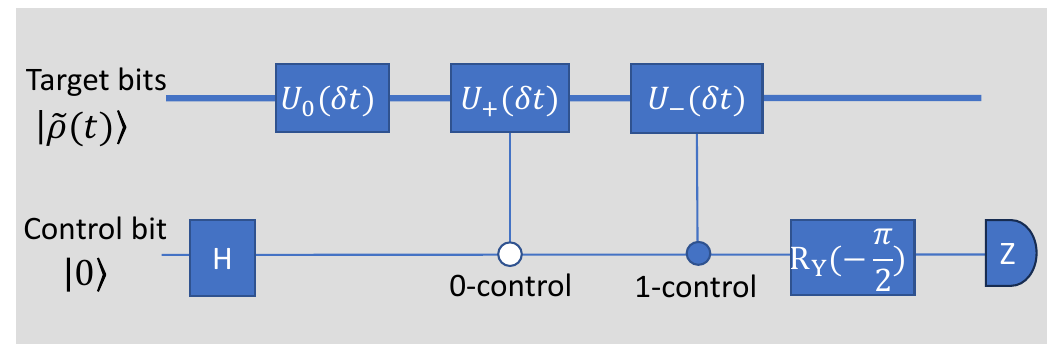}
  \caption{Illustration of quantum circuit for the temporal propagation of DQME-SQ within a single time step by utilizing the LCU algorithm.
}\label{fig2}
\end{figure}

\section{Representing DQME-SQ by quantum circuits}\label{thsec4}
%


The DQME-SQ offers a notable advantage: its formulation embraces a time-local structure within a compact Hilbert space.
This unique feature enables easy encoding into quantum circuits, distinguishing it from HEOM. 
While the computational complexity remains comparable to HEOM when implemented with classical algorithms, its excellent compatibility with quantum algorithms makes DQME-SQ a preferred choice for simulating complex non-Markovian OQSs.

To simulate the non-Markovian open quantum dynamics governed by the DQME-SQ using quantum circuits, the first key step is encoding the RDT.
For bosonic environments,
the RDT $\tilde{\rho}$ is mapped onto a pure qubit state, $\lvert \tilde{\rho} \rangle$, as
\begin{align}\label{encode}
 \tilde\rho \longmapsto |\tilde\rho \ra=\frac{1}{Z}\sum_{ij{\bf n}}P_{ij}^{\bf n}\,|i\ra\otimes |j\ra \otimes |\bf n\ra.
\end{align}
Here, the summation is over all the possible configurations of the dissipaton-embedded system, where $|{\bf n}\ra=|n_1\ra\otimes |n_2\ra\cdots\otimes |n_K\ra$ represents a dissipaton configuration. 
Note that for fermionic environments, $|{\bf n}\ra$ is replaced by paired dissipaton Fock states [cf.\,\Eq{s27}]. 

When implementing the DQME-SQ approach, it is necessary to truncate the dissipaton configurations. This truncation is particularly important in the case of bosonic environments, where the occupation number for a bosonic dissipaton level can, in principle, take on any value.
In the context of digital quantum simulations, the truncation of dissipaton configurations effectively reduces the dimension of the RDT, which in turn alleviates the computational burden associated with the DQME-SQ implementation.
The truncation of dissipaton configurations in the DQME-SQ approach is, in essence, equivalent to the truncation of the HEOM formalism. The latter has been extensively studied in the literature, and a variety of truncation schemes have been developed \cite{Tan06082001,Zha15214112,Han18234108,Zha21905,Din22224107,Zha23014106}.  
These established truncation schemes typically allow for rapid convergence of the results. However, in practice, it is still desirable to conduct careful convergence tests to ensure that the truncation errors are negligibly small.

The dissipaton configurations are encoded onto the qubits using a standard binary code.
The reduced density operator of the original system under investigation is given by $\rho_{\tS}(t)=\la \bm 0|\ti\rho(t)$, and its trace is preserved during the propagation of the DQME-SQ.
For the representation of the RDT on qubits, a renormalization procedure is required due to the normalization of the qubits' state.
The global normalization coefficient $Z$ is determined by the probability conservation condition $\sum_i P_{ii}^{\bf 0}=1$. 
It is important to note that this renormalization procedure does not cause additional computational burden or introduce numerical errors during the quantum simulation; see Appendix \ref{thapp2} for further discussions.

The temporal evolution of $\tilde{\rho}(t)$ is then simulated by propagating the following equation of motion,
\begin{align}
  |\tilde \rho(t)\ra=e^{-i\Lambda t} |\tilde \rho(0)\ra,
\end{align}
where $\Lambda$ is a non-Hermitian dynamical generator that represents the right-hand side of \Eq{dqme-sq} or \Eq{fermion}. 
It reads
\begin{align}\label{lambda}
  \Lambda&=({H}_{\tS}\otimes I_{\tS}-I_{\tS} \otimes H_{\tS}^{\rm T})\otimes I_{\D}\!-i I_{\tS}\otimes I_{\tS}\otimes \sum_k\!\gamma_k \hat b_k^+ \hat b_k^-
\nl & \quad
   \!+\!({\hat Q}\otimes I_{\tS}-I_{\tS}\otimes \hat{Q}^{\rm T})\otimes\sum_k \zeta_k (\hat b_k^- + \hat b_k^+)
   \nl
   &\quad 
   +i({\hat Q}\otimes I_{\tS}+I_{\tS}\otimes \hat{Q}^{\rm T})\otimes\sum_k \xi_k \hat b_k^+.
 \end{align}
%
%
%
%
Here, $I_{\tS}$ is the identity operator of the system Hilbert space ${\cal H}_{\tS}$,  $ I_{\D}$ is identity operator of the dissipaton Hilbert space ${\cal H}_{\D}$,
and $\hat O^{\rm T}$ denotes the transpose of any operator $\hat O$.
We then represent the state vector $|\ti\rho\ra$ on the qubits by using the standard binary code.
Equations (\ref{encode}) and (\ref{lambda}) constitute the  encoding protocol of the bosonic DQME-SQ.

The fermionic counterpart of \Eq{encode} reads
\begin{align}\label{s27}
\tilde{\rho} \longmapsto |\tilde\rho\ra = \frac{1}{Z}\sum_{\mu \nu{\bf mn} } P_{\mu \nu}^{\bf mn} |\mu\ra \otimes | \nu \ra \otimes|{{\bf m}}\ra \otimes|{\bf n} \ra.
\end{align}
Here,  $| \mu\ra, | \nu\ra \in \mathcal{H}_{\tS}$ are the system states.
The dynamical generator of \Eq{fermion} can be similarly constructed as
\begin{align}\label{s2856}
  &\Lambda= ({H}_{\tS}\otimes I_{\tS} -I_{\tS}\otimes H_{\tS}^{\rm T})\otimes I_{\D} \otimes I_{\D}  
  \nl & 
-iI_{\tS}\otimes I_{\tS}\otimes\Big(\sum_k \gamma^-_k \hat{N}_k \otimes {I}_{\D} +{I}_{\D}\otimes \sum_k \gamma^+_k \hat N_k\Big)
  \nl & 
+\sum_k\!\Big[\zeta^-_k \big(\hat c^+\otimes I_{\tS}  \otimes  \hat b_k^-\otimes {I}_{\D}-I_{\tS}\otimes \hat { c}^{-} \otimes  \hat b_k^-\otimes {I}_{\D}\big)
\nl & 
+\zeta^+_k\big(\hat c^-\!\otimes I_{\tS}  \otimes  {I}_{\D}\otimes \hat b_k^- - I_{\tS}\otimes \hat { c}^{+} \!\otimes   {I}_{\D}\otimes \hat b_k^-\big)
  \nl  &  
  +\xi^{+}_k  \hat c^{+}\otimes I_{\tS} \otimes {I}_{\D}\otimes \hat{ b}_k^{+}-\xi^{+\ast}_k  I_{\tS}\otimes  \hat{c}^{+}\otimes \hat b_k^+\otimes {I}_{\D}
    \nl  & 
  -\xi^{-}_k  \hat c^{-}\otimes I_{\tS} \otimes \hat b_k^+\otimes {I}_{\D} +\xi^{-\ast}_k  I_{\tS}\otimes \hat{ c}^{- }\otimes {I}_{\D}\otimes \hat{ b}_k^{+}\Big].
  \end{align}
The two parts, $|\mu\ra\otimes|\bf{m}\ra$ and  $| \nu\ra \otimes|{{\bf n}}\ra$ in $|\tilde \rho\ra$ of \Eq{s27}, are then encoded separately with  different sets of Jordan-Wigner codes.
Equations (\ref{s27}) and (\ref{s2856}) constitute the  encoding protocol of the fermionic DQME-SQ.

The temporal propagation of $|\tilde \rho(t)\ra$ can be realized
with various quantum algorithms, including the linear combination of unitaries (LCU) \cite{Lon06825,Sch22023216}, quantum imaginary time evolution \cite{Kam22010320,Mot20205}, quantum singular value transformation 
\cite{Low17010501,Low19163,Mar21040203,Sur231002}, time-dependent variational principle \cite{Sch21270503,Sch22023216,Hea21013182,Wan234851}, and others 
\cite{Lam23Arxiv2310_12539,Zha24054101,Sur231002}.
Here, we utilize the LCU propagation \cite{Sch22023216} for the demonstration.
%
Within a small time step $\delta t$, the propagator $e^{-i{ \Lambda}\delta t}$ is approximately decomposed as
$
e^{-i{ \Lambda}\delta t}
\approx U_{1}(\delta t)U_{0}(\delta t)
$.
Here, $U_0(\delta t)\equiv e^{-i\Lambda_0\delta t}$ governed by a Hermitian generator, $\Lambda_0\equiv (\Lambda+\Lambda^{\dg})/2$, is a unitary propagator, while $U_1(\delta t)\equiv e^{-i\Lambda_1\delta t}$ governed by an anti-Hermitian generator, $\Lambda_1\equiv (\Lambda-\Lambda^{\dg})/2$, is non-unitary.
The latter is further approximated by 
\cite{Sch21270503}
\begin{equation} \label{eps}
 U_1(\delta t) \approx
\frac{1}{2\epsilon}[U^{\epsilon}_{+}(\delta t)+U^{\epsilon}_{-}(\delta t)],
\end{equation}
where $U^{\epsilon}_{\pm}(\delta t)\equiv \pm ie^{\mp i\epsilon(I-i\Lambda_1\delta t)}$ is unitary 
and $\epsilon$ is a real parameter that takes a sufficiently small value.

Shown in  \Fig{fig2} is the designed quantum circuit of DQME-SQ comprising the 
 target bits to store the RDT state as well as a control bit for propagation.
The input of the circuit is $|\tilde \rho(t)\ra \otimes |0\ra$.
From the left to the right in the circuit, we first act a Hadamard gate (H) on the control bit, followed by the $U_{0}(\delta t)$ gate on the target bits.
Subsequently, the states go through the 0-control and 1-control  gates of $U^{\epsilon}_{\pm}(\delta t)$ successively. 
After acting an ${\rm R}_{\rm Y}(-{\pi}/{2})$ gate on the control bit, we obtain
$2\epsilon|\ti\rho(t+\delta t)\ra\otimes|0\ra+|{\rm unwanted}\ra\otimes|1\ra$.
Finally, by performing an Z-measurement on the control bit, we project out the unwanted state and obtain the target state $|\ti\rho(t+\delta t)\ra$ with a probability depending on $\epsilon$.
Once obtained, the wanted state will serve as the initial state for the next propagation step.

By using the Trotter decomposition, the unitary operators $U_{0}(\delta t)$ and $U_{\pm}^{\epsilon}(\delta t)$ can be split into a series of simpler quantum logic units. This facilitates the practical implementation of the DQME-SQ theory.
In contrast to classical algorithms, whose computational costs grow exponentially as the system size or the complexity of non-Markovianity increases \cite{Tan20020901}, the LCU-Trotter algorithm exhibits highly favorable scalability.  
Specifically, both the number of qubits and the circuit depth grow linearly with the increase in the  degrees of freedom of the dissipatons. This favorable scalability makes the DQME-SQ theory, implemented with the LCU-Trotter algorithm, as a promising approach to efficiently simulate complex non-Markovian open quantum dynamics. 

It should be mentioned that the unitary decomposition approximation in \Eq{eps} will incur a systematic error on the order of $O(\epsilon^2)$ \cite{Sch22023216}. 
For practical applications on real quantum computers in the noisy intermediate-scale quantum (NISQ) regimes and beyond, these errors and measurement costs from the unitary decomposition approximations can be significant \cite{Sch21270503, Sch22023216}.
A sufficiently large $\epsilon$ is needed to resolve the matrices representing the quantum gates from the device noise,
while a small $\epsilon$ is needed to obtain an accurate expansion of \Eq{eps}.
To address this conflict, techniques such as Richardson's extrapolation can be employed to improve the epsilon error and generate results with higher-order accuracy \cite{Sch21270503, Sch22023216}.
Additionally, amplitude amplification technology can be used to reduce the measurement costs by amplifying the probability of detecting the target state \cite{Bra0253, Kwo21062438,Ber, Sur231002}.  After considering both error control and amplitude amplification, the circuit depth of the LCU-Trotter algorithm grows polynomially as the degrees of freedom of dissipaton increases.  The  detailed analysis of complexity and error is presented in Appendix \ref{analy}. Therefore, the errors and measurement costs induced by the unitary decomposition approximation in \Eq{eps} are thus controllable and
expected to be gradually alleviated as quantum computing technologies advance.

\begin{figure}
\includegraphics[width=0.9\columnwidth]{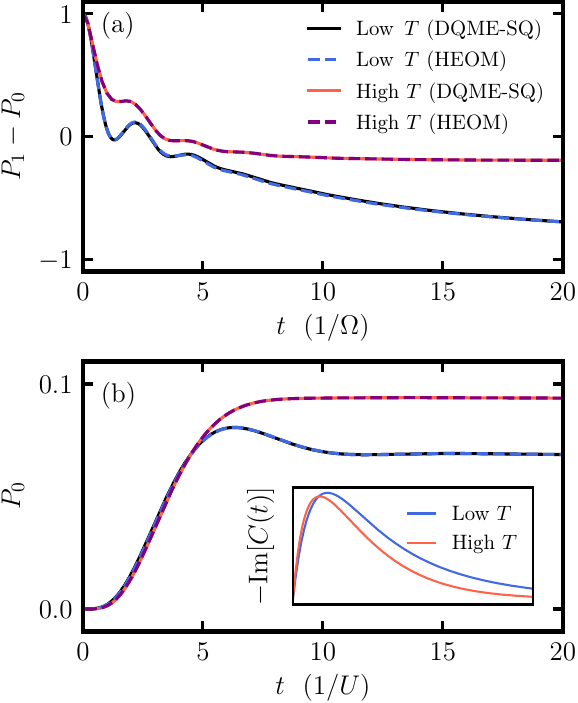}
  \caption{
(a) Simulated time evolution of  $P_1-P_0$, the population difference between two states, in a spin-boson model,  by using the quantum algorithm for the DQME-SQ theory, compared to low and high temperature results via the classical propagation of HEOM; 
(b) Simulated time evolution  of the ground-state population $P_0$ for a single-impurity Anderson model, with inset depicting the imaginary part of the correlation function. Detailed parameters are provided in the Appendix \ref{thapp3}.
} 
  \label{fig3}
\end{figure}

\section{Numerical demonstration through digital quantum simulation}\label{thsec5}

To demonstrate the feasibility of the DQME-SQ approach, we employ the LCU-Trotter algorithm to propagate the quantum dissipative dynamics for two exemplary models: the spin-boson model for the bosonic case, and the single-impurity Anderson model for the fermionic case.
For this purpose, we utilize the Aer simulator of Qiskit \cite{Qiskit} to propagate the DQME-SQ. Within each time step, we directly extract the target state, thus bypassing the need for extensive probabilistic measurements that would be required on actual quantum computing platforms.

The spin-boson model, with $H_{\tS}=\Omega\hat \sigma_z +V\hat \sigma_x$, is one of the most commonly used models that has broad applications in modern science.
Here, $\Omega$ and $V$ are the energy difference and transfer coupling between two local states. 
On the other hand, the single-impurity Anderson model, with $H_{\tS}  = E_0 (\hat n_{\uparrow} + \hat n_{\downarrow}) + U \hat n_{\uparrow} \hat n_{\downarrow}$, is widely used in condensed matter physics to investigate the strongly correlated transition metal impurities in metals.
Here, $E_0$ is the impurity level energy and $U$ denotes the electron-electron interaction energy. 
For bosonic and fermionic environments, the non-Markovian memory associated with the temperature is 
manifested in the real and imaginary parts of the hybridization correlation function, respectively.
In the fermionic case, as shown in the inset of \Fig{fig3}(b), ${\rm Im}[C(t)]$ exhibits a longer extension in the time domain at lower temperatures. This indicates a more pronounced non-Markovian characteristics, which is anticipated to have a more significant impact on the dissipative dynamics of the system.

As shown in \Fig{fig3}, the simulated time evolution results
obtained from DQME-SQ and HEOM exhibit perfect agreement, where the HEOM method utilizes classical computation algorithms.
Notably, \Fig{fig3} demonstrates that non-Markovian effects are more pronounced at lower temperatures in both bosonic and fermionic environments.
With all other parameters unchanged, a lower temperature leads to more pronounced oscillations in the short-time regime and slower relaxation in the long-time regime, due to longer environmental memory.

Simulations of more complex OQSs are demonstrated in the Appendix \ref{thapp3}. These additional examples further exhibit the exactness and universality of the DQME-SQ theory.
Due to limited computational resources at our disposal, the digital quantum simulations presented above are constrained to minimal models.
When implementing the DQME-SQ approach on realistic NISQ devices, the errors and measurement costs arising from the unitary decomposition approximations will need to be carefully considered and addressed.
Techniques such as those proposed in Refs.~\cite{Sch21270503,Sch22023216} can be employed to mitigate these issues.
However, as quantum computing technologies continue to advance significantly in the future, the utilized quantum algorithm, including the LCU-Trotter method, will become universally feasible for simulating more complex OQSs.

\section{Summary and perspective}\label{thsum}

Towards the quantum simulation of non-Markovian open quantum dynamics, this paper presents the DQME-SQ theory, a universal and exact theory that enables quantum simulation of dissipative dynamics in both bosonic and fermionic environments.
It captures the full non-Markovian nature of open quantum systems, provided that the environments follow Gaussian statistics.
By translating the non-Markovian memory content into a compact description of statistical quasi-particles (dissipatons), the DQME-SQ theory provides a physical interpretation of OQSs that is particularly suited for quantum computation, especially in fermionic environments. 
This framework provides extensive universality and flexibility for advancing quantum simulation in OQSs, enhancing studies of nonequilibrium dynamics and thermodynamics by elucidating the roles of non-Markovian influences in complex OQSs \cite{Cle101155,Soa14825,Coh15266802,Har201184}.
Moreover, it is also intriguing to explore the connection between the DQME-SQ and other recently developed methods, such as those based on the ensemble of trajectories \cite{Hea19022109} and the density matrix purification \cite{Sch22_arXiv_2207_07112}. These methods may share some useful techniques that could be beneficial for the quantum simulation of open quantum dynamics.

\begin{acknowledgments}
Support from the Ministry of Science and Technology
of China (Grant No.\ 2021YFA1200103), the National Natural Science Foundation of China (Grant Nos.\ 22425301, 22393912, 22321003, 22103073,  22173088, and 22373091), 
the Strategic Priority Research Program of Chinese Academy of Sciences (Grant No.\ XDB0450101), and the Innovation Program for Quantum Science and Technology
(Grant No.\ 2021ZD0303301) is gratefully acknowledged. The authors are indebted to Yu~Su, Daochi~Zhang,  and Zi-Fan Zhu for their invaluable help.
\end{acknowledgments}

\appendix 
\section{Relationship between bosonic DQME-SQ and pseudomode theories}\label{thapp1}
In this appendix, we will illustrate the underlying relationship between the DQME-SQ theory and the pseudomode theory. 
The pseudomode theory requires the coefficients $\{\eta_k\}$ in \Eq{exp} to be real-valued, as implied by Eq.\,(18) in Ref.\,\cite{Tam18030402} or Eq.\,(35) in the Supplemental Information of Ref.\,\cite{Lam193721}.
In contrast, the DQME-SQ theory is generally applicable with complex-valued $\{\eta_k\}$ and $\{\gamma_k\}$ in \Eq{exp}. This is a notable distinction between the two approaches.
However, in circumstances where both $\{\eta_k\}$ and $\{\gamma_k\}$  in \Eq{exp}  are real numbers, a clear relationship can be established between the DQME-SQ theory and the Lindblad-form pseudomode equation.

The  pseudomode Lindblad-type equation reads \cite{Tam18030402}
\begin{align}\label{psm_boson}
\!\!\!\!\dot{\rho}_{\rm p}(t)&= \!-i\big[H_{\tS},\rho_{\rm p}\big]\!-i\sum_k \zeta_k\Big[(\hat a^-_k+\hat a^+_k)\hat{Q},\rho_{\rm p}\Big]
 \nl & \quad
 +\sum_k\!\gamma_k \Big[2\hat a^-_k \rho_{\rm p} \hat a^+_k - \hat a^+_k \hat a^-_k \rho_{\rm p}\! -\rho_{\rm p} \hat a^+_k \hat a^-_k\Big].
\end{align}
Here, $\zeta_k$ is  defined before  \Eq{barrho}. $\{\hat a_k^{\pm}\}$ are the  creation and annihilation operators acting on the $k$th pseudomode, with
$\{|n_k\ra_{\rm p}\}$ being the associated Fock state. They follow the standard bosonic commutation algebra $[\hat a^-_k,\hat 
a^+_{k'}]=\delta_{kk'}$ and $[\hat a^\pm_k,\hat 
 a^\pm_{k'}]=0$.
 We aim to establish an explicit relationship [cf.\,\Eq{relation} in below] between 
$\rho_{\rm p}(t)\in \mathcal{H}_{\tS}\otimes \mathcal{H}_{\tS}^* \otimes {\cal H}_{\D}\otimes {\cal H}_{\D}^{*}$ (subscript p stands for pseudomode) in the pseudomode Lindblad equation of \Eq{psm_boson}  and  $\ti \rho(t)\in \mathcal{H}_{\tS}\otimes \mathcal{H}_{\tS}^* \otimes {\cal H}_{\D}$ 
in the DQME-SQ of \Eq{dqme-sq}, 
 under the condition $\{\xi_k=0\}$, which is a direct consequence of $\{\eta_k\}$ and $\{\gamma_k\}$ being real numbers. 
With this established relationship [cf.\,\Eq{relation} in below], 
we can retrieve the DQME-SQ from the pseudomode Lindblad equation, 
provided that  $\{\eta_k\}$ and $\{\gamma_k\}$ are real-valued. 

The explicit relationship reads
\begin{align}\label{relation}
\tilde{\rho}=\sum_{\bf n} \bigg[\big(\prod_k n_k !\big)^{-\frac{1}{2}} \,{\rm tr}_{\D}\Big({\mathcal N}\Big[\prod_k (\hat{a}^+_k + \hat{a}^-_k)^{n_k} \Big] \rho_{\rm p}\Big)\bigg] \otimes |{\bf n}\ra.
\end{align}
In \Eq{relation}, ${\mathcal N}$ denotes the normal ordering which places all creation operators to the left of all annihilation operators, 
 ${\rm tr}_{\D}$  denotes the partial trace over the pseudomode space $ {\cal H}_{\D}\otimes {\cal H}_{\D}^{*}$, and $\sum_{\bf n}$ represents the sum over all ordered set of indexes, ${\bf n}\equiv \{n_1\cdots n_K\}$.
By taking the time-derivative of $\ti \rho(t)$ on the left-hand-side of \Eq{relation}, 
\begin{align}\label{s15}
\dot{\tilde{\rho}}=\!\sum_{\bf n}\!\bigg\{\big(\prod_k n_k !\big)^{-\frac{1}{2}} \,{\rm tr}_{\D}\Big[{\mathcal N}\Big(\prod_k (\hat{a}^+_k + \hat{a}^-_k)^{n_k} \Big) \dot\rho_{\rm p}\Big]\bigg\} \otimes |{\bf n}\ra.
\end{align}
The substitution of pseudomode Lindblad equation \Eq{psm_boson} into \Eq{s15} leads to
\begin{equation}\label{dqme-sqB-easy}
\dot{\tilde \rho}=-i[H_{\tS},\tilde\rho]-\sum_k \gamma_k \hat  b_k^{\dg}\hat b_k\tilde \rho 
-i\sum_k \zeta_k(\hat  b_k^+ +\hat  b_k^-)[\hat{Q},\tilde\rho].
\end{equation}
This is the DQME-SQ \Eq{dqme-sq} with $\xi_k=0$ for all $k$.
The detailed derivation will be provided below. The one-to-one correspondence between the relevant quantities is illustrated in Table~\ref{demo-table}.

\begin{center}
\begin{table}[t]
    \caption{One-to-one correspondence between relevant quantities in \Eq{psm_boson} and \Eq{dqme-sqB-easy} }
\label{demo-table}
\def\arraystretch{2.2}
    \begin{tabular}{c|c} 
        \hline
    In \Eq{psm_boson} & In \Eq{dqme-sqB-easy} \\
\hhline{=|=}
 $-i[H_{\tS},\rho_{\rm p}]$ & $-i[H_{\tS},\ti\rho]$ \\
 \hline
 $\gamma_k \big(2\hat a^-_k \rho_{\rm p} \hat a^+_k - \hat a^+_k \hat a^-_k \rho_{\rm p} -\rho_{\rm p} \hat a^+_k \hat a^-_k\big)$&$-\gamma_k \hat  b_k^+ \hat 
 b_k^- \tilde \rho$
 \\
  \hline
 $-i\zeta_k\big[(\hat a^-_k+\hat a^+_k)\hat{Q},\rho_{\rm p}\big]$&$-i \zeta_k(\hat  b_k^+ +\hat  b_k^-)[\hat{Q},\tilde\rho]$
 \\
  \hline
    \end{tabular} 
    \end{table}
\end{center}

We now provide a three-step proof for the one-to-one correspondences listed in Table~\ref{demo-table}. First, we relate the pseudomode Lindblad equation of \Eq{psm_boson}  to the equation of motion for the intermediate variables $ \rho_{\bf{u, v}}(t)$ [\Eq{s19_19} in below]. Second, we make a direct connection between \Eq{s19_19} to \Eq{DEOM2} (with $\xi_k=0$ for all $k$).
Third, we validate the connection constructed in the above second step, and thus
retrieve the DQME-SQ of Eq.\eqref{dqme-sqB-easy} directly from \Eq{DEOM2}.

\paragraph*{Step 1: Relate \Eq{psm_boson} to \Eq{s19_19} for $ \rho_{\bf{u, v}}(t)$.} For later use, we define 
\be 
 \rho_{\bf{u, v}}(t)\equiv {\rm tr}_{\D}\bigg[\prod_k (\hat a^+_k)^{v_k}(\hat a^-_k)^{u_k} \rho_{\rm p}(t)\bigg].
\ee
Based on the pseudomode Lindblad equation of \Eq{psm_boson},  we obtain the equation of motion for $\rho_{\bf{u, v}}(t)$ as
\begin{align} \label{s19_19}
\dot\rho_{\bf{u, v}}(t)=&-i[H_{\tS},\rho_{\bf{u, v}}]-\sum_{k}\gamma_k  \Big(u_k\rho_{{\bf u}, {\bf v}}+v_k\rho_{{\bf u}, {\bf v}}\Big)
\nl &
-i\sum_{k} \zeta_k\Big[\hat{Q},\rho_{{\bf u}_k^{+}, {\bf v}}+\rho_{{\bf u}, {\bf v}_k^{+}}\Big]
\nl &
-i \sum_{k}\zeta_k\Big(u_k\hat Q \rho_{{\bf u}_k^{-}, {\bf v}}- v_k\rho_{{\bf u}, {\bf v}_k^{-}}\hat Q\Big).
\end{align}

\begin{center}
\begin{table}[t]
    \caption{One-to-one correspondence between relevant quantities in \Eq{psm_boson}  and \Eq{s19_19} }
\label{demo-table2}
\def\arraystretch{2.2}
    \begin{tabular}{c|c} 
        \hline
    In \Eq{psm_boson} & In \Eq{s19_19} \\
\hhline{=|=}
 $-i[H_{\tS},\rho_{\rm p}]$ & $-i[H_{\tS},\rho_{\bf{u, v}}]$\quad 
 \\[5pt]
 \hline
 \\[-18pt]
 \makecell{
 $\gamma_k \big(2\hat a^-_k \rho_{\rm p} \hat a^+_k - \hat a^+_k \hat a^-_k \rho_{\rm p}$ \quad 
 \\
 $- \rho_{\rm p} \hat a^+_k \hat a^-_k\big)$}&$-\gamma_k (u_k+v_k)\rho_{{\bf u}, {\bf v}}$
 \\ [8pt]
 \hline
 \\ [-18pt]
 $-i\zeta_k\big[(\hat a^-_k+\hat a^+_k)\hat{Q},\rho_{\rm p}\big]$&
\quad \makecell{$-i \zeta_k\big[\hat{Q},\rho_{{\bf u}_k^{+}, {\bf v}}+\rho_{{\bf u}, {\bf v}_k^{+}}\big]$\\  $-i \zeta_k\big(u_k\hat Q \rho_{{\bf u}_k^{-}, {\bf v}}- v_k\rho_{{\bf u}, {\bf v}_k^{-}}\hat Q\big)$}
 \\ [11pt]
  \hline
    \end{tabular} 
    \end{table}
\end{center}
The one-to-one correspondence between relevant quantities is illustrated in Table \ref{demo-table2}. The following relations are used
\begin{align*}
&{\rm tr}_{\D}\bigg[\Big(\prod_j (\hat a^+_j)^{v_j}(\hat a^-_j)^{u_j}\Big) \hat a^-_k\rho_{\rm p} \hat a^+_k\bigg]
\nl &\quad 
={\rm tr}_{\D}\bigg[\Big(\prod_j (\hat a^+_j)^{v_j+\delta_{jk}}(\hat a^-_j)^{u_j+\delta_{jk}}\Big)\rho_{\rm p} \bigg]
=\rho_{{\bf u}_k^{+}, {\bf v}_k^{+}},
\\
&{\rm tr}_{\D}\bigg[\Big(\prod_j (\hat a^+_j)^{v_j}(\hat a^-_j)^{u_j}\Big) \hat a^+_k \hat a^-_k\rho_{\rm p}\bigg]
=\rho_{{\bf u}_k^{+}, {\bf v}_k^{+}}+u_k\rho_{{\bf u}, {\bf v}},
\\
&{\rm tr}_{\D}\bigg[\Big(\prod_j (\hat a^+_j)^{v_j}(\hat a^-_j)^{u_j}\Big) \rho_{\rm p}\hat a^+_k \hat a^-_k\bigg]
=\rho_{{\bf u}_k^{+}, {\bf v}_k^{+}}+v_k\rho_{{\bf u}, {\bf v}},
\\
&{\rm tr}_{\D}\bigg[\Big(\prod_j (\hat a^+_j)^{v_j}(\hat a^-_j)^{u_j}\Big) (\hat a^-_k+\hat a^+_k)\rho_{\rm p} \bigg]
\nl &\quad 
=\rho_{{\bf u}_k^{+}, {\bf v}}+\rho_{{\bf u}, {\bf v}_k^{+}}+u_k\rho_{{\bf u}_k^{-}, {\bf v}},
\\
&{\rm tr}_{\D}\bigg[\Big(\prod_j (\hat a^+_j)^{v_j}(\hat a^-_j)^{u_j}\Big) \rho_{\rm p} (\hat a^-_k+\hat a^+_k)\bigg]
\nl &\quad 
=\rho_{{\bf u}, {\bf v}_k^{+}}+\rho_{{\bf u}_k^{+}, {\bf v}}+v_k\rho_{{\bf u}, {\bf v}_k^{-}}.
\end{align*}
In deriving above relations, we have used the invariance of trace and the Wick theorem.

\paragraph*{Step 2: Abridge \Eq{s19_19} to \Eq{DEOM2}, and \Eq{DEOM2} directly yields the DQME-SQ.}
Next, by using the binomial expansion, one then recast \Eq{relation} into
\begin{align}\label{sss242}
\tilde{\rho}&=\sum_{\bf n} \bigg[\big(\prod_k n_k !\big)^{-\frac{1}{2}} \,{\rm tr}_{\D}\Big({\mathcal N}\Big[\prod_k (\hat{a}^+_k + \hat{a}^-_k)^{n_k} \Big] \rho_{\rm p} \Big)\bigg] \otimes |{\bf n}\ra
\nl &
=\sum_{\bf n}  \bigg\{\big(\prod_k n_k !\big)^{-\frac{1}{2}}\sum_{{\bf u}\leq {\bf n}}\,{\rm tr}_{\D}\bigg[\prod_k {n_k \choose u_k}(\hat a^+_k)^{u_k} 
\nl & \quad\quad\quad\quad
\times (\hat a^-_k)^{n_k-u_k}\rho_{\rm p} \bigg] \bigg\}\otimes |{\bf n}\ra
\nl &
=\sum_{\bf n}\!  \bigg\{(\prod_k n_k !\big)^{-\frac{1}{2}}\sum_{{\bf u}\leq {\bf n}}\,\prod_k {n_k \choose u_k}\rho_{\bf{u, n-u}}  \bigg\}\!\otimes\!|{\bf n}\ra.
\end{align}
Here, ${\bf u}\leq {\bf n}$ means the set that satisfies $u_k\leq n_k$ for all $k$, and the definition of $\rho_{\bf{u, v}}$ is used in the last equality.
Equation (\ref{sss242}) shows that the RDT $\tilde{\rho}$ is more compact than $\rho_p$ in the pseudomode theory. To retrieve the former from the latter, one can first trace out the pseudomode degrees of freedom (${\cal H}_{\D}\otimes{\cal H}_{\D}^{\ast}$), and then use the dissipaton configurations in ${\cal H}_{\D}$ as a basis to express the components.

By comparing \Eq{sss242} with \Eq{defB}, we recognize
the relation between the dynamical variables in \Eq{s19_19} and \Eq{DEOM2}
\be \label{25s}
\bar {\rho}_{\bf n}^{(n)}= \big(\prod_k n_k !\big)^{-\frac{1}{2}}\sum_{{\bf u}\leq {\bf n}}\,\prod_k {n_k \choose u_k}\rho_{\bf{u, n-u}}.
\ee
Therefore, there is only one last step left, and that is  to prove the right-hand-side of \Eq{25s} also satisfies \Eq{DEOM2}, since the latter directly giving rise to
the DQME-SQ of \Eq{dqme-sqB-easy}.

\paragraph*{Step 3: Prove the right-hand-side of \Eq{25s} satisfies \Eq{DEOM2}.} Taking the time derivative of \Eq{25s} and combining the outcome with \Eq{s19_19}, we obtain 
\begin{align}\label{sy1}
\dot{\bar {\rho}}_{\bf n}^{(n)}&=\big(\prod_k n_k !\big)^{-\frac{1}{2}}\sum_{{\bf u}\leq {\bf n}}\,\prod_k {n_k \choose u_k}\dot\rho_{\bf{u,n-u}}
\nl &\quad
=(-i)\times({\rm I})+\sum_j[({\rm II})_j-i\zeta_j\times({\rm III})_j],
\end{align}
where
\begin{align}\label{sy2}
({\rm I}) &= \big(\prod_k n_k !\big)^{-\frac{1}{2}}\sum_{{\bf u}\leq {\bf n}}\,\prod_k {n_k \choose u_k}[H_{\tS},\rho_{\bf{u,n-u}}]
\nl &
=[H_{\tS},\bar {\rho}_{\bf n}^{(n)}],
\\ \label{sy3}
({\rm II})_j &=\big(\prod_k n_k !\big)^{-\frac{1}{2}}\sum_{{\bf u}\leq {\bf n}}\,\prod_k {n_k \choose u_k}\Big[-\gamma_j  (u_j+n_j-u_j)
\nl & \quad \times
\rho_{\bf{u,n-u}}\Big]
=- {n_j}\gamma_j \bar{\rho}^{(n)}_{{\bf n}},
\\ \label{sy4}
({\rm III})_j &=\big(\!\prod_k n_k !\big)^{-\frac{1}{2}}\!\sum_{{\bf u}\leq {\bf n}}\prod_k \!{n_k \choose u_k}
\!\Big\{\Big[\hat{Q},\rho_{{\bf u}_j^{+}, {\bf n-u}}
\!+\!\rho_{{\bf u}, ({\bf n-u})_j^{+}}\!\Big]
\nl &\quad 
+\Big(u_j\hat Q \rho_{{\bf u}_j^{-}, {\bf n-u}}- (n_j-u_j)\rho_{{\bf u}, ({\bf n-u})_j^{-}}\hat Q\Big)\Big\}
\nl & 
=\hat{Q}^{\times}\Big[\sqrt{n_j +1}\bar{\rho}^{(n+1)}_{{\bf n}^+_j}+\sqrt{n_j}\bar{\rho}^{(n-1)}_{{\bf n}^-_j} \Big].
\end{align}
In deriving \Eq{sy4}, we have used the following identities regarding combinatorics
\begin{align*}
  &\sum^{N}_{n=0} {N \choose n} \Big(x^{n+1}y^{N-n}+x^{n}y^{N-n+1}\Big)
  \nl &  
  =\sum^{N+1}_{m}{N+1 \choose m } x^m y^{N+1-m} =(x+y)^{N+1},
\end{align*}
and
\begin{align*}
(N-n){N \choose n}=N {N-1 \choose n}, \ \ {N \choose n}={N \choose N-n}.
\end{align*} 
Combing \Eqs{sy1}--(\ref{sy4}), we then obtain
\begin{align}
\dot{\bar{\rho}}^{(n)}_{\bf{n}}=&-i[H_{\tS},\bar{\rho}^{(n)}_{{\bf n}}] -\sum_k {n_k}\gamma_k \bar{\rho}^{(n)}_{{\bf n}}-i\sum_{k}\zeta_k \hat{Q}^{\times}
\nl & 
\Big[\sqrt{n_k +1}\bar{\rho}^{(n+1)}_{{\bf n}^+_k}+\sqrt{n_k}\bar{\rho}^{(n-1)}_{{\bf n}^-_k} \Big],  \label{eqn:deom3}
\end{align}
which is exactly \Eq{DEOM2} with $\{\xi_k=0\}$. 
Therefore, we have verified that, under the circumstances that $\{\eta_k\}$ and $\{\gamma_k\}$ are real numbers (and hence the condition $\{\xi_k=0\}$ is satisfied), the DQME-SQ recovers the pseudomode Lindblad equation exactly.


\section{Trace-preserving property of reduced density operator in DQME-SQ}\label{thapp2}

According to \Eq{defB}, the reduced density operator of the original system under investigation, $\rho_{\tS}(t)$, can be obtained by projecting the RDT of the dissipaton-embedded system onto the dissipaton vacuum state $\la {\bf 0}|$:
\begin{equation}
\rho_{\tS}(t)=\la {\bf 0}| \tilde{\rho}(t).    
\end{equation}
Consequently, the DQME-SQ of \Eq{dqme-sq} directly gives rise to 
\begin{equation}\label{1stt}
\dot{\rho}_{\tS}(t)=-i[H_{\tS},\rho_{\tS}]-i\sum_k \zeta_k[\hat{Q},\bar \rho_{k}^{(1)}],
\end{equation}
where $\bar \rho_{k}^{(1)}= \la {\bf 0}_{k}^{+}|\tilde\rho$ is an operator in the space of the system. Evidently, \Eq{1stt} preserves the trace of $\rho_{\tS}(t)$ at any time $t$, since the trace of any commutator is zero.
Therefore, the DQME-SQ can be directly solved in the time domain using any classical propagator without the need to renormalize the RDT at each time step.

When a quantum algorithm is used, the RDT $\ti \rho$ of the dissipaton-embedded system   needs to be encoded onto qubits states, which are inherently normalized.
Upon encoding $\ti \rho$ onto the qubit states, we have
\be\label{encoding56}
\tilde\rho \longmapsto |\tilde\rho \ra=\sum_{ij{\bf n}}T_{ij}^{\bf n}\,|i\ra\otimes |j\ra \otimes |\bf n\ra.
\ee
Since the qubit states $\vert i \ra$, $\vert j \ra$ and $\vert n \ra$ are inherently normalized, we must introduce the coefficients $T_{ij}^{\bf n}$ to ensure the vector state $\vert\tilde\rho \ra$ is also normalized. Specifically,  we require that $ \la \tilde\rho \vert \tilde\rho\ra = \sum_{ij{\bf n}} |T_{ij}^{\bf n}|^2=1$. However, this normalization condition generally means that ${\rm tr}_{\tS} (\rho_{\tS}) = \sum_i \la i \vert \otimes \la i \vert \otimes \la \bm 0 \vert \tilde\rho \ra = \sum_{i} T_{ii}^{\bf 0}\neq 1$. 
To preserve the trace of the reduced density operator and ensure ${\rm tr}_{\tS}(\rho_{\tS}) = 1$, we introduce the coefficients $P_{ij}^{\bf n}\propto T_{ij}^{\bf n}$, where $P_{ij}^{\bf 0}= \la i \vert \otimes \la j \vert \otimes \la \bm 0 \vert \tilde\rho \ra$ and ${\rm tr}_{\tS} (\rho_{\tS}) = \sum_{i} P_{ii}^{\bf 0}=1$, as described in \Eq{encode} of the main text.

It is important to note that the procedure for renormalizing the reduced density operator $\rho_{\tS}$ is not an inherent part of the DQME-SQ theory itself.
Rather, this renormalization is only required when a quantum algorithm is implemented to solve the DQME-SQ dynamics.
The same renormalization procedure is also necessary for quantum algorithms that solve other quantum dynamical equations, as long as a density operator needs to be encoded onto qubit states. For instance, Eq.\,(5) in Ref.\,\cite{Sch22023216} 
involves the encoding of so-called ``vectorized density matrix'' in the context of a Lindblad equation. 
Moreover, the renormalization procedure can be done only once after the entire evolution process is complete. Consequently, it does not impose any additional computational burden or introduce numerical errors during the quantum simulation process. Details on the quantum encoding protocols are provided in \Sec{thsec4}.

\section{Detailed analysis of error and  complexity}
\label{analy}
In this appendix, we will present more detailed analysis of error and complexity for DQME-SQ-based LCU-Trotter algorithm. 

The LCU-Trotter method exhibits an error of order $O(\epsilon^2) + O(\lambda K {\Delta}t^2)$ for evolution over a time step ${\Delta}t$. Here, $K$ represents the number of dissipaton modes, and $\lambda$ characterizes the system-environment coupling strength, with the dimension of $[T]^2$. The goal is to control the error within the precision, $\text{Err}$, during the evolution from time $0$ to $t$. Given that dissipative dynamics include decay terms that suppress divergence, it is assumed that the errors in each time step during the evolution are linearly additive. Thus, the total error is approximately $t O(\epsilon^2/{\Delta}t) + t O(\lambda K {\Delta}t) \sim \text{Err}$. This leads to the requirement that the step length be constrained as ${\Delta}t \sim \text{Err}/(t \lambda K)$, and the expansion coefficient as $\epsilon \sim \text{Err}/(t \lambda^{1/2} K^{1/2})$. Under this error control, as the number of dissipaton modes $K$ increases, the step size ${\Delta}t$ decreases. The circuit depth  grows polynomially ($\sim O(\lambda^{1/2} t^2 \text{Err}^{-1} K^2)$)  as the number of dissipaton modes increasing.

After a single step of evolution, the amplitude of the target state $|\tilde{\rho}\rangle \otimes |0\rangle$ is $2\epsilon$, so the probability of measuring the target state is only $4 \epsilon^2$ after performing a Z-measurement on the auxiliary qubit. Since the target state $|\tilde{\rho}\rangle \otimes |0\rangle$ is orthogonal to the unwanted state $|\text{unwanted}\rangle \otimes |1\rangle$, this issue can be addressed using existing amplitude amplification techniques \cite{Bra0253, Ber14283, Kwo21062438, Sur231002} to increase the measurement probability to the order of 1. Amplitude amplification, which shares its origins with the Grover search algorithm, works by applying multiple rotations to the superposition of the target state and its orthogonal state, thus enhancing the target state’s amplitude and improving the measurement probability. The number of rotations depends on $\epsilon$, scaling as $O(\epsilon^{-1})$. The circuit depth for each rotation operation is proportional to the LCU-Trotter circuit depth, which is $O(K)$ \cite{Ber14283}.
In conclusion, after considering both error control and amplitude amplification, as the number of dissipaton modes $K$ increases, the circuit depth of the LCU-Trotter algorithm grows polynomially ($\sim O(\lambda^{3/2} t^3 \text{Err}^{-2} K^{5/2})$).
There is still room for further optimization in this regard. In the future, we also aim to combine the DQME-SQ framework with more advanced quantum algorithms to simulate complex open quantum systems more efficiently and accurately.

\section{Details on the numerical implementation of DQME-SQ}\label{thapp3}

\subsubsection*{1. Details on the numerical examples}

For the spin-boson model, we choose the Brownian oscillator spectral density for the environment,
\be 
J(\omega)=2\lambda \omega^2_0\zeta\omega/[(\omega^2-\omega_{0}^2)^2+\omega^2\zeta^2].
\ee
Other parameters include  $T=\Omega/2$,  $\lambda=0.4\,\Omega$, and  $V=\zeta=\omega_0=\Omega$. 
The number of dissipaton levels is $K=3$ and  the maximum allowed occupation number for each bosonic dissipaton level is $N_{\rm max}=3$.
%
%
%
%
For the propagation of $|\tilde \rho\ra$,  we adopt the time step of $\delta t=0.01\,\Omega^{-1}$ and the expansion coefficient $\epsilon=0.05$ in the linear combination of unitaries (LCU) algorithm.
The high temperature case of $T=5\,\Omega$ is also examined for comparison.
Initially, the system is in the spin-up state $|1\ra$. The explicit exponential decomposition reads (in units of $\Omega^2$ and $\Omega$ for $\{\eta_k\}$ and $\{\gamma_k\}$, respectively) 
\begin{align*}
C_{\text{Low T}}(t)&=(0.497+0.082i)e^{-(0.500+0.866i)t}
\nl & \quad
+(0.035-0.082i)e^{-(0.500-0.866i)t}
\nl & \quad
+(-0.032)e^{-3.873t}
\end{align*}
and
\begin{align*}
C_{\text{High T}}(t)&=(2.231+1.155i)e^{-(0.500+0.866i)t}
\nl &\quad
+(1.769-1.155i)e^{-(0.500-0.866i)t}.
\end{align*}

For the single-impurity Anderson model,  we consider a Lorentzian form for the environment spectral density as 
\be 
J(\w) = \Gamma W^2/(\w^2 + W^2). 
\ee
The energetic parameters involved in the model take the following values:
$E_0 = -U/2$, $W = U$, $\Gamma=U/8$ and $\beta U=4$ (high temperature) and $8$ (low temperature). 
The number of dissipaton levels is chosen to be $6$ ($K=3$ for each spin orbital), 
which approaches the upper limit of our computational capacity.
In the simulation of fermionic OQSs, all the dissipaton configurations are explicitly included without any truncation.
For the propagation of $|\tilde \rho\ra$, we  adopt the time step of $\delta t=0.01\,U^{-1}$ and the expansion coefficient $\epsilon=0.005$ in the LCU algorithm.
The system is initially in the double occupancy state.
The explicit exponential decomposition reads 
\begin{align*}
C_{\text{Low T}}^\pm(t)&=(0.062-0.038i)e^{-t}
+(-0.037i)e^{-0.393t}
\nl & \quad
+
(0.075i)e^{-1.630t}\end{align*}
and
\begin{align*}
C_{\text{High T}}^\pm(t)&=(0.062+0.138i)e^{-t}+(-0.164i)e^{-0.786t}
\nl & \quad
+(0.026i)e^{-3.261t}
\end{align*}
in units of $U^2$ and $U$ for the coefficients and exponents, respectively.
The codes are available at  
\MYhref[blue]{https://github.com/icct-ustc/dqme-sq}{https://github.com/icct-ustc/dqme-sq}.

\begin{figure}[t]
\includegraphics[width=\columnwidth]{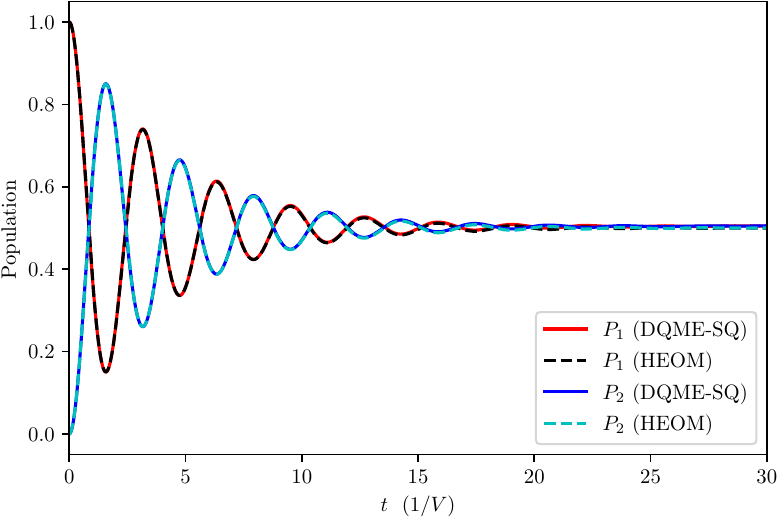}
\caption{Time evolution of the excitonic dimer system. The results are
obtained  by using the quantum algorithm for the DQME-SQ, compared to results via the classical propagation of HEOM. Plotted are  the populations in two excitons.} \label{fig_sm_bose}
\end{figure}

\begin{figure}[t]
\includegraphics[width=\columnwidth]{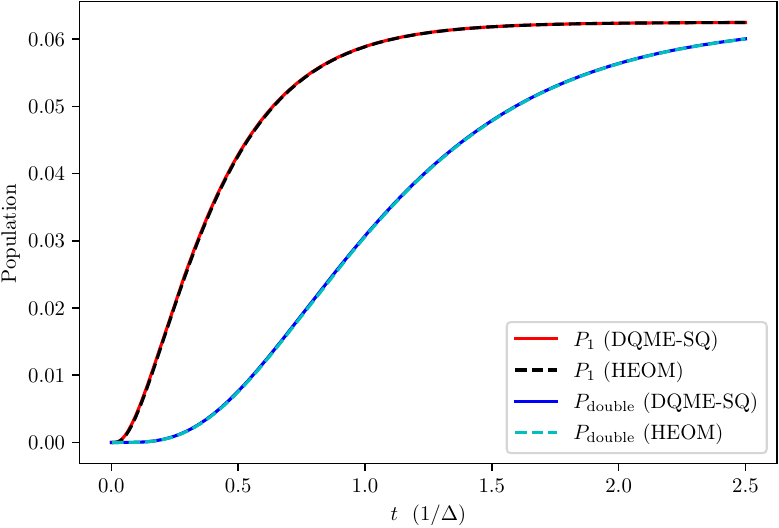}
  \caption{
  Time evolution of the double-impurity Anderson model, calculated from both DQME-SQ and HEOM. Plotted are the populations, exemplified with $P_1$ and $P_{\text{double}}$. 
} 
\label{fig_sm_fermion}
\end{figure}

\subsubsection*{2. Examples on more complex OQSs}

We present numerical examples on more complex OQSs.
These additional examples involve more than one degree of freedom in the system that is coupled to the environment. They serve to further validate the exactness and universality of the DQME-SQ theory we have developed.

\paragraph*{Bosonic case.} For the bosonic scenario, consider an excitonic dimer system with Hamiltonian:
\be
  H_{\tS}=\sum_{u=1}^{2}\varepsilon_{u}\hat{B}_{u}^{\dg}\hat{B}_{u}+V(\hat{B}_{1}^{\dg}\hat{B}_{2}+\hat{B}_{2}^{\dg}\hat{B}_{1}).
\ee
Here, the double excitation state is neglected, $\hat{B}_{u}\equiv |0\ra\la u|$ ($\hat{B}_{u}^{\dag}\equiv |u\ra\la 0|$) are the excitonic annihilation (creation) operators, and $\{\varepsilon_u\}$ and $V$ are the on-site energies and interstate coupling, respectively. The excitonic system is coupled to the harmonic bath environment through
\be
  H_{\tS\E}=\sum_{u=1}^{2}\hat{B}_{u}^{\dg}\hat{B}_{u}\hat{F}_u.
\ee
The bath hybridization spectral density takes the Drude form
\be\label{Drude}
  J_{11}(\w)=J_{22}(\w)=\frac{2\lambda\gamma\w}{\w^2+\gamma^2},
\ee
and off-diagonal hybridization is neglected, $J_{12}=J_{21}=0$. Take $V$ as the reference unit of energy and set $T=\varepsilon_1=\varepsilon_2=V$.
The Drude parameters are set to be $\lambda=0.5\,V$ and $\gamma=5\,V$.
 The number of dissipaton states is chosen to be $4$
($K=2$ for each exciton mode), using the classical HEOM and quantum DEOM-SQ algorithms. 
We set $N_{\rm max}=3$ to be the maximum allowed occupation number for each bosonic dissipaton level.
 As illustrated in \Fig{fig_sm_bose}, the results of these two methods appear to match perfectly, where $P_u\equiv \la \hat{B}_{u}^{\dg}\hat{B}_{u}\ra$ with $u=1,2$.


\paragraph*{Fermionic case.} For the fermionic scenario, consider the Anderson model with two magnetic impurities coupled to a single electron reservoir.
The Hamiltonian  reads
\begin{align}
H_{\T} = H_{\tS} + \!\sum_{k}\epsilon_{ks}\hat d^\dagger_{ks}\hat d_{ks} + \!\sum_{us}(\hat F^\dagger_{us}\hat a_{us} + \hat a_{us}^\dagger\hat F_{us}),\!
\end{align}
where the system Hamiltonian reads 
\begin{align}
    H_{\tS} &= \sum_{u=1,2}\varepsilon_u\hat n_u + U \sum_{u=1,2}\hat n_{u\uparrow}\hat n_{u\downarrow} + U_{\rm C}\hat n_1\hat n_2 
    \nl & \quad
    + t\sum_{s} (\hat a_{1s}^\dagger\hat a_{2s} + \hat a_{2s}^\dagger\hat a_{1s}),
\end{align}
with $\hat n_{u} \equiv \hat n_{u\uparrow} + \hat n_{u\downarrow}$ and $\hat n_{us} \equiv \hat a^\dagger_{us}\hat a_{us}$. We also set $\varepsilon_1 = \varepsilon_2 = -(U + 2U_{\rm C})/2$, with the spectral density
\begin{align}
J_{uvs}^\sigma(\omega) = \delta_{uv}\frac{\Delta W^2}{\omega^2 + W^2}.
\end{align}
%
The related parameters are given by $W = 50\Delta$, $U = U_{\rm C} = 12\Delta$, $t = 10\Delta$, and temperature $T=5\Delta$.
The number of dissipaton levels is chosen to be $4$
($K=1$ for each spin orbital), using the classical HEOM and DQME-SQ algorithms. 
All the dissipaton configurations are included without any truncation.
As illustrated in \Fig{fig_sm_fermion}, the results from these two methods match perfectly, where $P_1\equiv \la1_{1\uparrow} 1_{1\downarrow}0_{2\uparrow}  0_{2\downarrow} |\rho_{\tS}|1_{1\uparrow} 1_{1\downarrow}0_{2\uparrow}  0_{2\downarrow}\ra$
 and $P_{\text{double}}
 \equiv \la 1_{1\uparrow} 1_{1\downarrow}1_{2\uparrow}  1_{2\downarrow} |\rho_{\tS}|1_{1\uparrow} 1_{1\downarrow}1_{2\uparrow}  1_{2\downarrow}\ra$.


\end{document}